%% file: tnfcolloq.tex
\documentclass[rmp,twocolumn,aps]{revtex4-1}

\usepackage{graphicx}
\usepackage{verbatim}
\usepackage{nicefrac}
\usepackage{amsmath,amssymb,bm}

\newcommand{\beq}{\begin{equation}}
\newcommand{\eeq}{\end{equation}}
\newcommand{\beqa}{\begin{eqnarray}}
\newcommand{\eeqa}{\end{eqnarray}}

\newcommand{\Mhi}{M_{\rm high}}
\newcommand{\Mlo}{M_{\rm low}}
\newcommand{\Vlowk}{V_{{\rm low}\,k}}

\begin{document}

\title{Three-body forces: From cold atoms to nuclei}

\author{Hans-Werner Hammer}
\email{hammer@hiskp.uni-bonn.de}
\affiliation{Helmholtz-Institut f\"ur Strahlen- und Kernphysik (Theorie) \\ 
and Bethe Center for Theoretical Physics, Universit\"at Bonn,
D-53115 Bonn, Germany}

\author{Andreas Nogga}
\email{a.nogga@fz-juelich.de}
\affiliation{Institut f\"ur Kernphysik, Institute for Advanced Simulation
and J\"ulich Center for Hadron Physics,
Forschungszentrum J\"ulich, D-52425 J\"ulich, Germany}

\author{Achim Schwenk}
\email{schwenk@physik.tu-darmstadt.de}
\affiliation{Institut f\"ur Kernphysik, Technische Universit\"at Darmstadt, 
D-64289 Darmstadt, Germany \\ 
and ExtreMe Matter Institute EMMI, GSI Helmholtzzentrum f\"ur 
Schwerionenforschung GmbH, D-64291 Darmstadt, Germany}


\begin{abstract} 
\input{abstract}
\end{abstract}


\maketitle

\tableofcontents

\input{introduction}
\input{definition}
\input{universal}
\input{chpt}
\input{manybody}
\input{other}
\input{outlook}

\bibliographystyle{apsrmp}
\bibliography{lit-database/lit}    

\end{document}

%% file: abstract.tex
It is often assumed that few- and many-body systems can be accurately
described by considering only pairwise two-body interactions of the
constituents. We illustrate that three- and higher-body forces enter
naturally in effective field theories and are especially prominent in
strongly interacting quantum systems. We focus on three-body forces
and discuss examples from atomic and nuclear physics. In particular,
we highlight the importance and the challenges of three-nucleon forces
for nuclear structure and reactions, including applications to
astrophysics and fundamental symmetries.

%% file: introduction.tex
\section{Introduction}

In this colloquium, we discuss recent advances, challenges, and
perspectives of three-body forces in nuclear physics and related
areas. We start with a brief overview of the history of the subject.

The simplest non-relativistic system in which three-body forces can
appear is the three-body system. The study of the three-body problem
has a long history in physics. The gravitational problem of the
earth-moon-sun system was first considered by Newton
(\cite{Newton:1687aa}). It was a central topic in mathematical physics
from the mid 1700's to the early 1900's. In gravity, only two-body
interactions between point masses are present. However, three-body
tidal forces arise if extended objects, such as planets, are treated
as point particles.

Three-body forces also play an important role in quantum mechanics and
the quantum many-body problem. If they are not already present at a
fundamental level, three- and higher-body forces appear in effective
theories or in practical calculations, where the degrees of freedom
and the Hilbert space have to be restricted. Typically, there is a
hierarchy of these forces and two-body forces provide the main
contribution with three- and higher-body forces giving smaller and
smaller corrections.

\begin{figure}[t]
\begin{center}
\includegraphics[width=2.5cm,clip=]{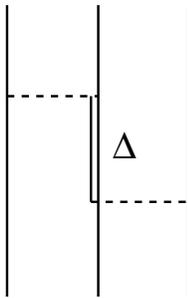}
\end{center}
\caption{Three-nucleon force arising from virtual excitation of 
a $\Delta(1232)$ degree of freedom. Solid (dashed) lines indicate
nucleons (pions).}
\label{fig:FM}
\end{figure}

A well known example of such a three-body force in nuclear physics is
the Fujita-Miyazawa three-nucleon force (\cite{Fujita:1957zz}).  Its
main contribution arises from the virtual excitation of a
$\Delta$(1232) resonance in processes involving three nucleons
interacting via pion exchanges illustrated in Fig.~\ref{fig:FM}. In
atomic, molecular and optical physics and quantum chemistry,
non-additive forces arise if Born-Oppenheimer potentials are
calculated by integrating out the electronic degrees of freedom
(\cite{Kaplan:1994aa}). In the theory of strong interactions, Quantum
Chromodynamics (QCD), three-body forces arise already at a fundamental
level as the three-gluon vertex induces an interaction between three
quarks.

Our discussion here is guided by effective field theory ideas. The
concept of {\it resolution} plays a key role in this context. 
In a scattering experiment, a particle beam 
with de Broglie wavelength $\lambda$ can only probe structures 
at a scale $R \gtrsim \lambda$. Similarly, in a general process with typical 
momentum scale $\mu$ only physics at momenta  $p \lesssim \mu$
(or, equivalently, distances $R \gtrsim 1/\mu$)
is resolved. Effective theories and the renormalization group provide a 
method to use this observation for quantitative calculations. The 
resolution scale in an effective theory is controlled by the momentum
cutoff $\Lambda$. Physics at momentum scales larger than the cutoff
is excluded from the effective theory and encoded in effective
couplings, so-called low-energy constants.
These constants and the
relative size of two- and higher-body forces turn out to be
resolution dependent. If one starts with two-body forces only at high
resolution, many-body forces will appear naturally as the resolution
scale is lowered. These induced many-body forces capture the contributions
of successive two-body interactions which are separated by a distance
below the resolution scale.

The natural size of many-body forces is determined by the underlying
scales of the theory and the external momentum scale $\mu$. 
Effective field
theories provide a convenient and systematic scheme to construct and
estimate the size of many-body forces. It is one aim of this
colloquium to exemplify these features. For example, chiral effective
field theories in the Weinberg scheme have a clear hierarchy of
many-nucleon forces. The current state of the art is to include two-
and three-nucleon forces. In the future, however, the inclusion of
four-body forces may also be required to achieve the desired
accuracy.

The quantitative importance of three-body forces is well established
in light nuclei. Therefore, they should contribute significantly in
heavier nuclei as well. However, their inclusion in many-body
calculations is computationally challenging and has only become
feasible in recent years. We discuss three-nucleon forces at different
resolution scales and show that their inclusion is mandatory for
nuclear structure calculations. Whether this scheme breaks down for
heavy nuclei beyond a certain mass number is an open question, but at
present there are no indications of such a breakdown.

The outline of the colloquium is as follows: We start with a review of
the theoretical framework for three-body forces including an
illustration of their scheme dependence. In Section~\ref{sec:univ}, we
discuss the role of three-body forces in the universal regime of large
scattering-length systems and give examples from nuclear and cold atom
physics. This is followed by a discussion of three-nucleon forces in
chiral effective field theory in Section~\ref{sec:chpt}. The
application of such forces to many-body systems and their relation to
electroweak processes is presented in Sections~\ref{sec:manybody}
and~\ref{sec:other}. Finally, we give an outlook and discuss future
opportunities.

%% file: definition.tex
\section{Theoretical approaches to three-body forces:
definitions, dependence on scheme and framework}
\label{sec:def}

Effective field theory (EFT) provides a general approach to understand
the low-energy behavior of a physical system.  The underlying
principle was concisely formulated in \cite{Weinberg:1978kz}: The most
general Lagrangian consistent with all symmetries of the underlying
interaction will generate the most general S-matrix consistent with
these symmetries. If this idea is combined with a power counting
scheme that specifies which terms are required at a desired accuracy,
one obtains a predictive low-energy theory.  The expansion is usually
in powers of a low-momentum scale $\Mlo$, which can be the typical
external momentum, over a high-momentum scale $\Mhi$. To illustrate
this idea, consider a theory that is made of two particle species, a
light and a heavy one with $\Mlo \ll \Mhi$. We focus on soft processes
in which the energies and momenta are of the order of the light
particle mass (the so-called soft scale).  Under these conditions, the
short-distance physics related to the heavy-particle exchange cannot
be resolved.  However, it can be represented systematically by contact
interactions between light particles.  Consider heavy-particle
exchange between the light ones at momentum transfer $q^2 \ll
\Mhi^2$. The corresponding tree-level expression for the scattering
amplitude is simply $g^2r/(\Mhi^2 - q^2)$ with $g$ the heavy-light
coupling constant. It can be expanded in powers of $q^2/\Mhi^2$ as:
\beq
\frac{g^2}{\Mhi^2 - q^2} = \frac{g^2}{\Mhi^2} + 
\frac{g^2 \, q^2}{\Mhi^4} + \ldots \,.
\label{eq:HeavyLight}
\eeq
This expansion can be represented in the EFT.  At low momentum
transfer $q^2$, the effects of the pole from the heavy-particle
exchange in Eq.~(\ref{eq:HeavyLight}) are captured by a series of
local momentum-dependent interaction terms reproducing the expansion
in Eq.~(\ref{eq:HeavyLight}) term by term. This idea is closely
related to the multipole expansion in classical electrodynamics and
the renormalization group (\cite{Wilson:1983dy}).

The interactions in EFTs are represented by operators ${\cal O}_i$
that are monomials in the quantum fields $\psi$ in the general interaction
Lagrangian,
\beq
{\cal L}_{\rm int} = \sum_i g_i \, {\cal O}_i \,.
\eeq
These operators can contain an arbitrary number of derivatives and/or
fields but must respect the symmetries of the underlying theory. The
derivatives are converted to momenta and 
generate the momentum dependence exemplified in
Eq.~(\ref{eq:HeavyLight}).  The coupling constants $g_i$ can be
ordered according to their importance at low energies from their
scaling with $q \sim \Mlo$ and $\Mhi$. Operators with a larger number of
derivatives or fields are usually suppressed.  This is the basis of the
power counting of the EFT.  

An illustrative example is given by the Lagrangian
\beq
{\cal L}_{\rm int} = \sum_{i=2}^N g_i \, (\psi^\dagger \psi)^i \,,
\eeq
which contains momentum-independent
two-, three-, ... up to $N$-body contact interactions
of a nonrelativistic field $\psi$. 
In a natural theory without any fine tuning of parameters, 
the dimensionful coupling constants $g_i$
scale with powers of the high-momentum or breakdown scale $\Mhi$.
Dimensional analysis requires that 
$g_i \sim (1/\Mhi)^{3i-5}$, such that
$N$-body interactions are suppressed by $(\Mlo/\Mhi)^{2(N-2)}$
compared to $N-1$ successive
two-body interactions (see, e.g., \cite{Hammer:2000xg}).
This type of scaling analysis 
is the basis of the suppression of many-body forces in the 
Weinberg scheme mentioned in the introduction.

The values of the coupling constants $g_i$ are determined completely
by on-energy-shell information, up to a well-defined truncation
error. The exact relation, however, is not unique and depends on the
renormalization scheme.  In the construction of the most general
Lagrangian, many-body forces arise naturally. These many-body forces
have to be determined from many-body data.

A fundamental theorem of quantum field theory states that physical
observables are independent of the choice of fields in a Lagrangian
(\cite{Haag:1958vt,Coleman:1969sm}).  Consequently, they are invariant
under redefinitions of the fields in the effective
Lagrangian. Off-shell amplitudes, however, change under field
redefinitions and thus are not observable.  In systems with more than
two nucleons, one can trade off-shell, two-body interactions for
many-body forces.

In a quantum mechanics framework, unitary transformations provide an
alternative formalism to field redefinitions and lead to the same
result (\cite{Polyzou:1990isi,Amghar:1995av}).  This explains how
two-body interactions related by unitary transformations can predict
different binding energies for the triton (\cite{Afnan:1973plb}) if
many-body forces are not consistently included.

\begin{figure}[t]
\begin{center}
\includegraphics[width=8cm,clip=]{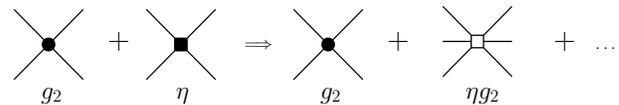}
\end{center}
\caption{Illustration of the field redefinition in Eq.~(\ref{eq:FD})
that trades an off-shell two-body interaction for a three-body
interaction.}
\label{fig:FD}
\end{figure}

We use a simple EFT model, to illustrate how 
field redefinitions can be used to shift strength from off-shell 
two-body interactions to on-shell three-body interactions
(\cite{Hammer:2000xg}):
\begin{equation}
{\cal L} = \psi^\dagger {\cal D} \psi - g_2 (\psi^\dagger \psi)^2
-\eta \left( \psi^\dagger (\psi^\dagger \psi) 
{\cal D}\psi +\psi^\dagger {\cal D}(\psi^\dagger \psi) \psi\right) \,,
\end{equation}
where ${\cal D} = i\partial_t +\vec \nabla^2/(2m)$ is the free Schr\"odinger
operator. The model has a two-body contact interaction with coupling
constant $g_{2}$ and an off-shell two-body contact interaction with 
coupling $\eta$ which we assume to be small. Now consider a field 
transformation
\begin{equation}
\label{eq:FD}
\psi \longrightarrow \left[1+\eta (\psi^\dagger \psi)\right]\psi\,,\quad
\psi^\dagger \longrightarrow \left[1+\eta (\psi^\dagger \psi)\right]\psi^\dagger\,.
\end{equation}
Performing this transformation and keeping all terms
of order $\eta$ we obtain a new Lagrangian:
\beq
{\cal L'} =  \psi^\dagger {\cal D} \psi - g_2 (\psi^\dagger \psi)^2
-4 \eta  g_2 (\psi^\dagger\psi )^3 + {\cal O}(\eta^2)\,,
\eeq
where the off-shell two-body interaction has been traded for a
three-body interaction. This is illustrated in Fig.~\ref{fig:FD}.
Off-shell interactions always contribute together with many-body
forces and only the sum of the two is meaningful.  It is thus not
possible to determine off-shell interactions from experiment.

On the other hand, if new interactions are generated from a given
two-body interaction by field redefinitions or unitary transformations
and only the two-body part is retained, many-body observables will
depend on the interaction even if they generate the same two-body
observables (see \cite{Furnstahl:2000we} for an explicit illustration
in the above model).  For example, nuclear matter binding curves will
depend on the off-shell part of the two-body interaction, generating
so-called ``Coester bands'' (\cite{Coester:1970ai}). In the 1970s, it
was proposed that comparisons of nuclear matter calculations could
help to determine the ``correct'' off-shell behavior of the two-body
interaction. From a modern perspective, it is clear that this is not
possible and the bands will disappear if the full transformed
Hamilitonian is used.

One might argue that it should be possible to find a suitable
representation of the theory where three-body forces vanish or are
very small.  This strategy could be used to minimize the computational
effort in many-body calculations. As discussed below, this is indeed
possible for the universal EFT but only at leading order (LO). In
theories with more complex operator structure and long-range
interactions such as the chiral EFT, however, it is doomed to fail
from the start.  The various operators contribute differently to
different observables and there is no optimal choice for removing the
contributions of three-body forces at the same time for all
observables.

\begin{figure}[t]
\begin{center}
\includegraphics[width=7.5cm,clip=]{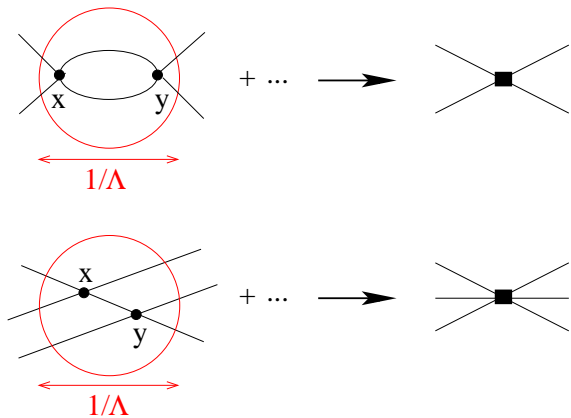}
\end{center}
\caption{Illustration for the resolution dependence of two- and
three-body interactions.}
\label{fig:highmom3nf}
\end{figure}

As illustrated above, the interaction strength may be shifted from
two- to many-body forces. This can also be seen by changing the
momentum cutoff $\Lambda$ in the regulators used in explicit
calculations.  Once the couplings $g_i$ of the effective Lagrangian,
the low-energy constants (LECs), have been adjusted to selected data,
predictions for other low-energy observables should be independent of
the choice for $\Lambda$. Obviously, this adjustment depends on
$\Lambda$ implying that also the interaction strengths of two- and
many-body interactions vary with the cutoff and are not unique.  The
idea is illustrated in Fig.~\ref{fig:highmom3nf}, where iterated
two-body interactions at short-distance scales of $\sim 1/\Lambda$ are
not resolved. Note that in practice the 3NFs generated in this way
cannot be disentangled from the 3NFs at an initial scale, which will
also have short-ranged (and other) contributions. It is therefore
model-dependent to distinguish such ''generated'' from ''genuine''
3NFs as is done often in the literature and, therefore, we will not
distinguish 3NF mechanisms in this colloquium.

For two-body forces, this can be implemented by renormalization group (RG) 
equations for the potential (\cite{Bogner:2006vp,Bogner:2009bt})
\begin{equation}
\label{eq:nnrg}
\frac{d}{d\Lambda } \, V_{\Lambda}(12) = F(V_{\Lambda},\Lambda) \,.
\end{equation}
The RG equation describes the evolution with $\Lambda$ of the matrix
elements $V_\Lambda(12)$ of the potential in momentum
space. The function $F$ is defined such that
the on-shell $T$-matrix is invariant under changes of the cutoff for
momenta below $\Lambda$.  This equation can be integrated from large
$\Lambda$ to lower cutoff scales.  By construction, all two-body
observables up to momenta of the order of the cutoff are
invariant. Beyond this, e.g., for processes involving external probes
and more particles, observables will depend on the cutoff. Complete RG
invariance is only achieved when many-body forces and many-body
currents are included. In principle, a similar RG equation for three-
and higher-body interactions can be formulated. This has been realized
in practice with the similarity RG (SRG) (\cite{Bogner:2006pc}) or by
taking EFT three-nucleon forces (3NFs) as a general low-momentum basis
(\cite{Nogga:2004ab}), where the LECs are adjusted to
few-nucleon data at the lower cutoffs. If the resolution scale
$\Lambda$ is not too low, the contributions of many-body forces
obtained in this way are of the size expected in EFT and small
compared to the two-nucleon (NN) force contributions, but they are
still quantitatively important in state-of-the-art computations.  The
variation of the cutoff then enables one to estimate contributions of
higher-body short-range interactions. This will be discussed in more
detail in Section~\ref{sec:chpt}.

While the RG evolution is already an interesting tool to estimate
many-body contributions to specific observables, it becomes even more
valuable in many-body calculations (see Section~\ref{sec:manybody}),
where RG transformations to lower resolution lead to greatly enhanced
convergence (\cite{Bogner:2009bt}).

%% file: universal.tex
\section{Universal aspects:
From cold atoms to low-energy reactions and halo nuclei}
\label{sec:univ}

As discussed in the previous section, the short-distance properties of
a physical system are not resolved in low-energy observables. If no
massless particles are present, all interactions appear short-ranged
at sufficiently low energy.  It is then possible to formulate an EFT
with contact interactions.

Particularly interesting is the case of strong interactions
characterized by a large scattering length $a$. Such systems are close
to the unitary limit of infinite scattering length.  It is obtained by
taking the range of the interaction to zero while keeping a 
two-body bound state fixed at zero energy.\footnote{In 
real physical systems the strict unitary 
limit of zero-range interactions can, of course, not be reached. 
In low-energy observables, however, the finite range $R$ is not resolved
and corrections are small (of the order of $R/a$ or $R \, k$).}
In this limit, the two-body scattering amplitude is scale invariant
and saturates the unitarity bound.  Formulated as a challenge to test
many-body methods (\cite{Bertsch:1999MBX}), this limit turned out to
be relevant for a variety of systems. It is historically interesting
to note that an approximation corresponding to the unitary limit was
already used in (\cite{Beth:1937zz}) to calculate the second virial
coefficient of a Fermi gas.  Ultracold atomic gases can be tuned to
the vicinity of the unitary limit using Feshbach resonances, while
neutron matter is close to this limit through a fine tuning in
nature. This gives rise to novel many-body phenomena, such as the
BEC-BCS crossover in ultracold atoms (\cite{Giorgini:2008zz}) and the
``perfect'' liquid observed in heavy-ion collisions
(\cite{Schafer:2009dj}).

Here, we use the unitary limit as a starting point for an EFT
expansion for strongly interacting quantum systems with short-range
interactions.  This universal EFT is applicable to any system close to
the unitary limit, i.e., any system with short-range interactions and
large scattering lengths.  Examples include halo states in nuclear
physics, ultracold atoms close to a Feshbach resonance, and hadronic
molecules in particle physics.  The breakdown scale $\Mhi$ of this
theory is set by the lowest energy degree-of-freedom not explicitly
included in the theory. In nuclear and particle physics, this is
typically given by one-pion exchange. In ultracold atoms, $\Mhi$ is
determined by the van der Waals interaction, but the details depend on
the system. The typical momentum scale of the theory is $\Mlo \sim 1/a
\sim k$.  For momenta $k$ of the order of the breakdown scale $\Mhi$
or above, the omitted short-range physics is resolved and has to be
treated explicitly.

The universal EFT exploits the appearance of a large scattering
length, independent of the mechanism generating it. Because the
dependence of observables on the scattering length is explicit, it
allows to unravel universal phenomena driven by the large scattering
length such as universal correlations of observables
(\cite{Phillips:1968isi}, \cite{Tjon:1975plb}), the Efimov effect
(\cite{Efimov:1970zz}), and limit-cycle physics
(\cite{Braaten:2003eu}, \cite{Mohr:2005pv}).  For reviews of
applications to the physics of ultracold atoms, see
\cite{Braaten:2004rn} and \cite{Platter:2009gz}.  The applications in
nuclear and particle physics were discussed in \cite{Epelbaum:2008ga}
and \cite{Hammer:2010kp}.

Three-body forces play an important role in the universal EFT and we
discuss their contribution in three- and higher-body systems in detail
below. In the simplest case of spinless bosons, the leading-order
Lagrangian can be written as:
\beq
{\cal L} = \psi^\dagger \Big(i\partial_t +\frac{\vec{\nabla}^2}{2m}\Big)
\psi -g_2 (\psi^\dagger \psi)^2 -g_3 (\psi^\dagger \psi)^3 +\ldots \,. 
\label{lagPsi}
\eeq
Extensions to more complicated systems are straightfoward.
The terms proportional to $g_2$ and $g_3$ correspond to
two- and three-body contact interactions. The dots represent 
higher-order terms suppressed by derivatives and/or more fields.

The renormalized values of the coupling constants $g_2$ and $g_3$
are matched to observables in the two- and three-body system.
In the two-body system, one typically takes the S-wave scattering length.
The exact relation between the coupling $g_2$ and the scattering length
depends on the renormalization scheme.
Because of this matching procedure, the EFT provides correlations 
between different observables based on the hierarchy of scales in the
system. Given one set of observables, another set can be predicted to
a certain accuracy. Depending on the experimental situation, these 
correlations can be applied in different ways. 

\begin{figure}[t]
\begin{center}
\includegraphics*[width=8cm,clip=]{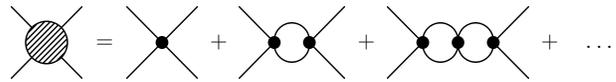}
\end{center}
\caption{The bubble diagrams with the contact interaction 
$g_2$ contributing to the two-body scattering amplitude.}
\label{fig:bubble}
\end{figure}

Since the scattering length is large, $a\sim 1/\Mlo$, the leading
contact interaction $g_2$ has to be resummed to all orders
(\cite{Kaplan:1998we,vanKolck:1998bw}).  The two-body scattering
amplitude is obtained by summing the bubble diagrams with the $g_2$
interaction shown in Fig.~\ref{fig:bubble}.  This summation gives the
exact solution of the Lippmann-Schwinger equation for the $g_2$
interaction and reproduces the leading term of the effective range
expansion.  Higher-order derivative interactions, which are not shown
explicitly in Eq.~(\ref{lagPsi}), generate higher-order terms in the
effective range expansion. Since these terms are set by $\Mhi$, their
contribution at low energies is suppressed by powers of $\Mlo /\Mhi$
and can be treated in perturbation theory. The first correction is
given by the S-wave effective range, $r_0 \sim 1/\Mhi$.

The S-wave scattering amplitude to next-to-leading 
order (NLO) then takes the form
\beq
T_2(k) = 
\frac{1}{-1/a-ik} \left[
1-\frac{r_0 k^2/2}{-1/a-ik}+\ldots \right]\,,
\label{eq:ereffects}
\eeq
where $k$ is the relative momentum of the particles and the dots
indicate corrections of order $(\Mlo /\Mhi)^2$ for typical momenta
$k\sim\Mlo$. If $a$ is large and positive, $T_2$ has a bound state
pole at $k=i/a$. This corresponds to a two-body bound state (dimer)
with binding energy $B_2=1/(2\mu a^2)$, where $\mu$ is the reduced
mass of the particles. As $a\to\infty$, this bound state approaches
the two-body threshold.

The universal EFT shows its full strength in the two-body sector when
external currents are considered. In contrast to other approaches, the
coupling to currents is straightforward and current conservation is
satisfied at each stage of the calculation. Gauge-invariant few-body
contact terms are generated naturally by writing the most general
effective Lagrangian. Applications to a variety of electroweak
processes in the two-nucleon sector have been carried out (see
\cite{Beane:2000fx}; \cite{Bedaque:2002mn} for more details).
Recently, these methods have also been applied to neutron-rich systems
and halo nuclei.  In \cite{Hammer:2011ye}, e.g., the electric
properties of the one-neutron halo nucleus $^{11}$Be were
investigated.  While this nucleus is nominally an 11-body system, the
properties of its ground and first excited state can be described in
the framework of the halo EFT
(\cite{Bertulani:2002sz,Bedaque:2003wa}). This EFT exploits the small
binding energy of these two states compared to the typical energy
scales of $^{10}$Be (binding and excitation energies). Thus, $^{11}$Be
can be treated as an effective two-body system of the $^{10}$Be core
and a neutron.  A similar strategy was applied to calculate the
radiative neutron capture on a $^7$Li core (\cite{Rupak:2011nk};
\cite{Fernando:2011ts}).

\begin{figure}[t]
\begin{center}
\includegraphics*[width=7cm,clip=]{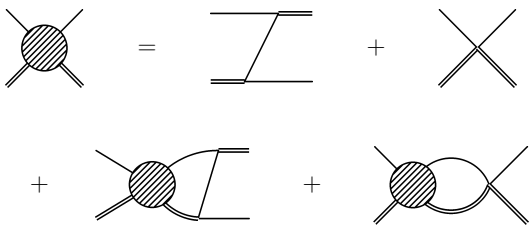}
\end{center}
\caption{The integral equation for the boson-dimer scattering amplitude.
The single (double) line indicates the boson (dimer)
propagator.}
\label{fig:ineq}
\end{figure}

We now proceed to the three-body system where the term proportional to
$g_3$ in Eq.~(\ref{lagPsi}) contributes. From naive dimensional
analysis one would conclude that the $g_3$ term is of higher order
(cf. discussion in Sec.~\ref{sec:def}).  This is indeed the case for
two-component fermions where the Pauli principle forbids three
fermions to be close together in an S-wave.  In general, however,
naive dimensional analysis fails for large scattering length
$a$. Again, we focus on the case of identical bosons which already
contains the main features of the problem.  The simplest three-body
process to be considered is the scattering of a boson and a dimer.
The integral equation for boson-dimer scattering is shown
schematically in Fig.~\ref{fig:ineq}. 
For total orbital angular momentum $L=0$, it takes the form:
\beqa
T_3(k, p; \, E) &=& \frac{16}{3 a} \, M (k, p; \, E) + \frac{4}{\pi} 
\int_0^\Lambda dq\, q^2\, 
T_3(k, q; \, E) \nonumber \\
&&\times \, \frac{ M (q, p; \, E)}{- 1/a + \sqrt{3 q^2/4 - m E - i \epsilon}} 
\,, 
\label{STM}
\eeqa
where the inhomogeneous term reads
\beq
M (k, p; \, E) = \frac{1}{2 k p} \ln 
\left( \frac{k^2 + k p + p^2 - m E}{k^2 - k p + p^2 - m E} \right)
+ \frac{H (\Lambda )}{\Lambda^2}\,,
\eeq
and a momentum cutoff $\Lambda$ has been introduced to regulate
the integral equation.
All other three-body observables can be extracted from the amplitude
$T_3$ taken in appropriate kinematics.
In Eq.~(\ref{STM}), $H$ determines  the strength of the three-body
interaction $g_3(\Lambda)=-4m\,g_2(\Lambda)^2 H(\Lambda)/\Lambda^2$.
The magnitude of the incoming (outgoing) relative momenta is
$k$ ($p$) and $E = 3 k^2/(4 m) - 1/(ma^2)$. 
The on-shell point corresponds to $k = p$ and 
the scattering phase shift can be obtained via
$k \cot \delta = 1/T_3(k, k; \, E)+ik$.

For $H=0$ and $\Lambda \to \infty$,  Eq.~(\ref{STM}) reduces to the
STM equation first derived by Skorniakov and Ter-Martirosian
(\cite{Skorniakov:1957aa}) which has 
no unique solution (\cite{Danilov:1961aa}).
The regularized equation has a unique solution for any given (finite) value 
of the cutoff $\Lambda$ but 
three-body observables show a strong dependence on the cutoff $\Lambda$.
Cutoff independence of the amplitude is restored by an appropriate 
``running'' of $H (\Lambda )$ which turns out to be 
a limit cycle (\cite{Bedaque:1998kg,Bedaque:1998km}):
\beq
H(\Lambda)\approx\frac{\cos[s_0 \ln(\Lambda/\Lambda_*)+\arctan s_0]}
{\cos[s_0 \ln(\Lambda/\Lambda_*)-\arctan s_0]}\,,
\label{eq:Heq}
\eeq
where $s_0\approx 1.00624$ is a transcendental number and 
$\Lambda_*$ is a dimensionful three-body parameter generated
by dimensional transmutation. Adjusting 
$\Lambda_*$ to a single three-body observable allows to determine all 
other low-energy properties of the three-body system. Note 
that the choice of the three-body parameter $\Lambda_*$ is not unique
and there are other definitions more directly related to 
experiment (\cite{Braaten:2004rn}).

\begin{figure}[t]
\begin{center}
\includegraphics[width=8cm,clip=]{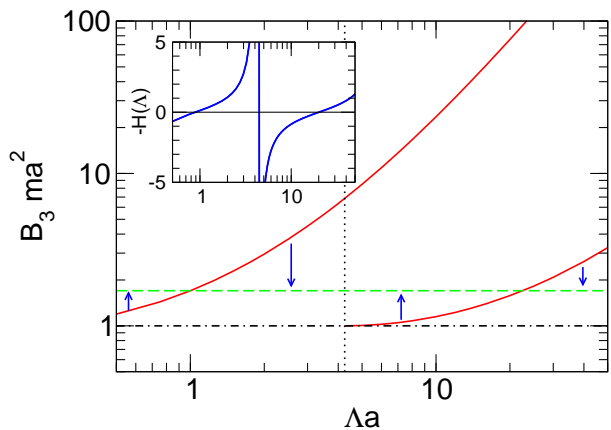}
\end{center}
\caption{(Color online) Unrenormalized three-body energies $B_3$ 
as a function of the momentum cutoff $\Lambda$ (solid lines).  
The dotted line indicates the cutoff where a new three-body state appears 
at the boson-dimer threshold (dash-dotted line). The dashed line 
shows a hypothetical renormalized energy. The inset shows the running
of the three-body force $g_3(\Lambda) \sim - H(\Lambda)$ with $\Lambda$.}
\label{fig:B3lambda}
\end{figure}

The physics of this renormalization procedure is illustrated in
Fig.~\ref{fig:B3lambda} where we show the unrenormalized three-body
binding energies $B_3$ in the case of positive scattering length as a
function of the cutoff $\Lambda$ (solid line).  As the cutoff is
increased, $B_3$ increases. At a certain cutoff (indicated by the
dotted line), a new bound state appears at the boson-dimer
threshold. This pattern repeats every time the cutoff increases by the
discrete scaling factor $\exp(\pi/s_0)$.  Now assume that we adopt the
renormalization condition that the shallowest state should have a
constant energy given by the dashed line. At small values of the
cutoff, we need an attractive three-body force to increase the binding
energy of the shallowest state as indicated by the arrow. As the
cutoff is increased further, the required attractive contribution
becomes smaller and around $\Lambda a =1.1$ a repulsive three-body
force is required (downward arrow).  Around $\Lambda a=4.25$, a new
three-body state appears at threshold and we cannot satisfy the
renormalization condition by keeping the first state at the required
energy anymore. The number of bound states has changed and there is a
new shallow state in the system. At this point the three-body force
turns from repulsive to attractive to move the new state to the
required energy.  The corresponding running of the three-body force
with the cutoff $\Lambda$ is shown in the inset.  After
renormalization, the first state is still present as a deep state with
large binding energy, but for threshold physics its presence can be
ignored. This pattern goes on further and further as the cutoff is
increased.

The three-body force in Eq.~(\ref{eq:Heq}) has exactly the right
behavior to implement the strategy from the previous paragraph.
Moreover, it breaks the scale invariance in the unitary limit, because
the three-body parameter $\Lambda_*$ now provides a scale. However,
due to the specific form of Eq.~(\ref{eq:Heq}), a discrete scale
invariance survives. Scaling transformations with the scaling factor
$\lambda_0=\exp(\pi/s_0)$ leave $H(\Lambda)$ and, consequently,
three-body observables invariant. This discrete scaling symmetry is
the signature of an RG limit cycle (\cite{Wilson:1970ag}).  In the
three-body bound-state spectrum it becomes manifest through the Efimov
effect: The appearance of a geometric spectrum of three-body bound
states (\cite{Efimov:1970zz}).

\begin{figure}[t]
\begin{center}
\includegraphics[width=6cm,clip=]{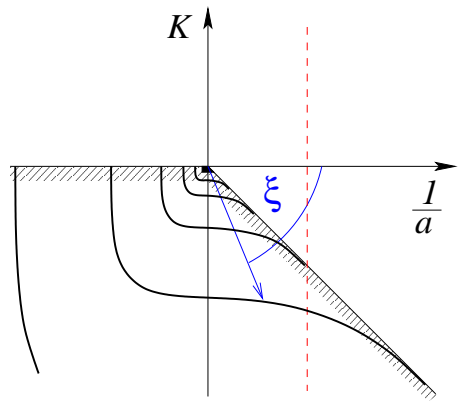}
\end{center}
\caption{(Color online) Illustration of the Efimov spectrum: The
energy variable $K={\rm sgn}(E)\sqrt{m|E|}$ is shown as a 
function of the inverse scattering length $1/a$. The solid lines
indicate the Efimov states while the hashed areas give the scattering 
thresholds. The dashed vertical line indicates a system with
fixed scattering length.}
\label{fig:efiplot}
\end{figure}

The Efimov spectrum is illustrated in Fig.~\ref{fig:efiplot}.  We show
the energy variable $K={\rm sign}(E)\sqrt{m|E|}$ as a function of the
inverse scattering length $1/a$. The hashed areas indicate the
three-atom ($a<0$) and atom-dimer thresholds ($a>0$) where the Efimov
states become unstable. The spectrum is invariant under the discrete
scaling transformations $K\to \lambda_0 K$ and $1/a \to
\lambda_0/a$. As a consequence, there is an accumulation of Efimov
three-body states at the origin.  The scaling symmetry relates Efimov
states along any ray with fixed angle $\xi$
(cf.~Fig.~\ref{fig:efiplot}). In general, these states correspond to
different scattering lengths.  A physical system with fixed scattering
length is illustrated by the vertical dashed line. For fixed $a$, the
discrete scaling symmetry is only manifest in the unitary limit
$1/a=0$.

The parameter $\Lambda_*$ can be used to set one of the three-body
energies. All other states then follow from the discrete scaling
symmetry.  This explains why one parameter is sufficient for
renormalization of the whole spectrum. The discrete scaling symmetry
predicts infinitely-deep three-body states. This is known as the
Thomas collapse (\cite{Thomas:1935zz}).  Physically relevant, however,
are only states with energies $|E|\ll \Mhi^2/m$. All deeper states are
ultraviolet artefacts of the effective theory and should be discarded.

\begin{figure*}[t]
\begin{center}
\includegraphics[width=16cm,clip=]{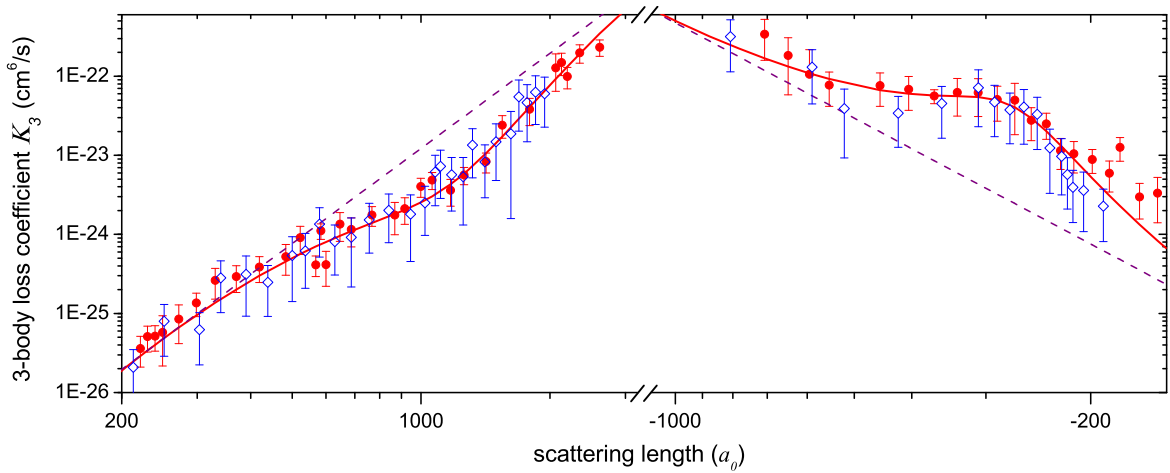}
\end{center}
\caption{(Color online) Three-body loss coefficient $K_{3}$ in a gas
of ultracold $^7$Li atoms as a function of scattering length (in
units of the Bohr radius $a_{0}$) for the $|m_{F}=1\rangle$ state
(red solid circles) and the $|m_{F}=0\rangle$ state (blue open
diamonds).  The solid lines represent fits to the universal EFT
prediction. The dashed lines represent the $K_3 \sim a^{4}$ upper
(lower) limit for $a>0$ ($a<0$). (Figure taken from \cite{Gross:2010aa}.)}
\label{fig:133Cs}
\end{figure*}

The discrete scale invariance also manifests itself in the
log-periodic dependence of scattering observables on the scattering
length. This scaling behavior has been confirmed in cold atom
experiments (\cite{Ferlaino:2010viw}). In such experiments, the
scattering length can be varied using Feshbach resonances.  The
scattering-length dependence of three-body recombination rates
provides indirect information on the Efimov spectrum. For negative
scattering length, the Efimov states hit the three-atom threshold,
$E=0$, for certain values of $a$ (cf.~Fig.~\ref{fig:efiplot}) and lead
to enhanced recombination rates. For positive scattering length, the
Efimov states become unstable already at the atom-dimer threshold,
$E=-1/(ma^2)$, but interference effects lead to minima and maxima in
the rate at $E=0$. Ideally one would like to see multiple
recombination features on each side of the Feshbach resonance. For
equal mass particles, this is not a simple task because of the large
scaling factor.  When effective-range effects are included
perturbatively as in Eq.~(\ref{eq:ereffects}), the discrete scale
invariance is softly broken, but the effects of the breaking on the
recombination rate can be calculated (\cite{Ji:2011qg}).

As an example, we show in Fig.~\ref{fig:133Cs} the three-body loss
coefficient $K_{3}$ in a gas of ultracold $^7$Li atoms measured by the
Khaykovich group (\cite{Gross:2010aa}) as a function of scattering
length (in units of Bohr radius $a_{0}$) for the $|m_{F}=1\rangle$
state (red solid circles) and the $|m_{F}=0\rangle$ state (blue open
diamonds).  The data show that the positions and widths of
recombination minima and Efimov resonances are identical for both
states, which indicates that the short-range physics is nuclear-spin
independent.  The solid lines give fits to the analytical expressions
of the universal EFT (\cite{Braaten:2004rn}) and reproduce the data
very well.

A more direct way to observe Efimov states is to populate these states
directly through radio frequency transitions.  This is difficult
because of their short lifetime and has only recently been achieved
for $^6$Li atoms~(\cite{Lompe:2010ads,Nakajima:2011ads}).

The integral equations for the three-nucleon problem are a
generalization of Eq.~(\ref{STM}).  (For their explicit form and
derivation, see \cite{Bedaque:2002yg}.)  The leading-order three-body
force is required in all channels where short distances are not
shielded by the angular momentum barrier and/or the Pauli principle.
For S-wave nucleon-deuteron scattering in the spin-quartet channel the
three-body force is of higher order,
and the spin-quartet scattering phases can therefore be predicted to
high precision from two-body data
(\cite{Bedaque:1997qi,Bedaque:1998mb}).  In the spin-doublet channel
there are two coupled channels but the renormalization is similar to
the three identical-boson case.  Thus, one needs a new parameter which
is not determined in the NN system in order to determine the (leading)
low-energy behavior of the three-nucleon system in this
channel. A comprehensive discussion of three-body force effects in the
three-nucleon system was given by \cite{Griesshammer:2005ga}.

The three-body parameter gives a natural explanation of universal
correlations between different three-body observables such as the
Phillips line: a correlation between the triton binding energy and the
spin-doublet neutron-deuteron scattering length
(\cite{Phillips:1968isi}).  These observables are calculated for
different two-body potentials that reproduce the NN scattering phase
shifts but the three-body parameter is not constrained by the
data. This generates a one-parameter correlation between different
three-body observables.  These correlations are driven by the large
scattering length and are independent of the mechanism responsible for
it.  As a consequence, they occur in atomic systems such as $^4$He
atoms as well (\cite{Braaten:2004rn}).  For an overview of this topic
see \cite{Epelbaum:2008ga} and \cite{Hammer:2010kp}.

\begin{figure}[t]
\begin{center}
\includegraphics[width=8cm,clip=]{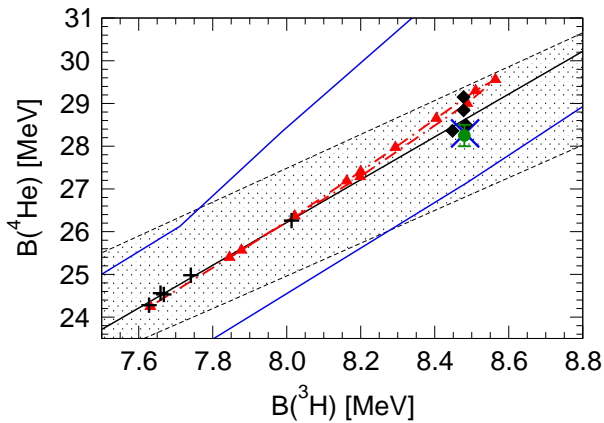}
\end{center}
\caption{(Color online) The Tjon line correlation between $B(^3{\rm H})$ and
$B(^4{\rm He})$. The experimental value is shown by the cross.
The shaded band gives the LO universal result (\cite{Platter:2004zs}),
while the outer solid lines include NLO corrections obtained 
using the resonating group method
(\cite{Kirscher:2009aj}). The pluses and diamonds show calculations using
phenomenological NN and NN+3N potentials, respectively 
(\cite{Nogga:2000uu}); the circle gives a chiral EFT result 
at N$^2$LO (\cite{Epelbaum:2002vt}); and the triangles are based
on an SRG-evolved N$^3$LO NN potential 
(\cite{Nogga:2004ab}; \cite{Hebeler:2010xb}).\label{fig:tjon}}
\end{figure}

The universal EFT has also been applied in the four-body sector. A
study of the cutoff dependence of the four-body binding energies
revealed that no four-body force is required for renormalization at
leading order (\cite{Platter:2004qn,Platter:2004zs}).  Thus, the
four-body force is a higher-order effect.  As a consequence, there are
also universal correlations in the four-body sector driven by the
large scattering length.  The prime example is the Tjon line
(\cite{Tjon:1975plb}): a correlation between the triton and
alpha-particle binding energies, $B(^3{\rm H})$ and $B(^4{\rm He})$.
Higher-order range corrections break the correlation and generate a
band.  In Fig.~\ref{fig:tjon}, we show this band together with
calculations using phenomenological NN potentials
(\cite{Nogga:2000uu}), a chiral NN potential at
next-to-next-to-leading order (N$^2$LO) 
(\cite{Epelbaum:2002vt}), SRG-evolved next-to-next-to-next-to-leading
order (N$^3$LO) 
NN potentials (\cite{Nogga:2004ab}; \cite{Hebeler:2010xb}), and the
experimental value.  All calculations with interactions that give a
large scattering length must lie within the band. Different
short-distance physics and/or cutoff dependence should only move the
results along the band. This can be observed explicitly in the results
for the SRG-evolved N$^3$LO NN potential indicated by the triangles in
Fig.~\ref{fig:tjon}, as well as in few-body calculations with
low-momentum interactions $\Vlowk$ (\cite{Nogga:2004ab}).

The absence of a four-body force at leading order also implies a
universal four-body spectrum.  In \cite{Hammer:2006ct} the dependence
of the four-body bound-state spectrum on the two-body scattering
length was investigated in detail and summarized in a generalized
Efimov plot for the four-body spectrum. In particular, it was found
that there are two four-body states tied to every Efimov trimer.  In a
subsequent study, \cite{vonStecher:2008nat} extended these
calculations to the four-particle threshold and confirmed the absence
of a four-body parameter for shallow four-body states.  Their
prediction of the resonance positions lead to the experimental
observation of universal tetramer states in ultracold caesium
(\cite{Ferlaino:2009zz}). This, in turn, has led to increased
theoretical activity in this area.  The sensitivity of tetramer
energies to a four-body scale was, for example, investigated by
\cite{Hadizadeh:2011qj}. Four-body recombination and other scattering
processes were calculated by Deltuva (see \cite{Deltuva:2012zy} and
references therein).

The bound-state properties of larger systems of bosons interacting
through short-range interactions were considered by
\cite{Hanna:2006aa}. Using Monte Carlo methods they showed that
universal correlations between binding energies can also be obtained.
Calculations for larger number of particles using a model that
incorporates the universal behavior of the three-body system were
carried out by \cite{vonStecher:2009qw}. These findings indicate that
there is at least one $N$-body state tied to each Efimov trimer and
numerical evidence was also found for a second excited 5-body
state. In a subsequent study (\cite{vonStecher:2011aa}), the energies
and structural properties of bosonic cluster states up to $N=6$ were
calculated for various two-body potentials. Besides the lowest cluster
states, which behave as bosonic droplets, cluster states bound weakly
to one or two atoms forming effective cluster-atom "dimers" and
cluster-atom-atom "trimers" were identified. For a related study in
the hyperspherical harmonic basis, see (\cite{Gattobigio:2012}). Thus
the prospects for observing universal physics in larger few-body
systems are excellent. Note that coherent multi-body interactions of
bosonic atoms have also been observed in a three-dimensional optical
lattice (\cite{Will:2010}).

Recently, a geometric spectrum of universal three-body states has also
been predicted for atoms with dipolar interactions
(\cite{Wang:2011aa}).  In this case, the structure of the interaction
is very similar to the nuclear tensor force generated by one-pion
exchange.  If the dipole moments of the atoms are aligned, the
interaction is attractive in a head-to-tail configuration of the atoms
and repulsive side-by-side, like for dipole magnets. If the dipole
moments are anti-aligned, the interaction is opposite, repulsive and
attractive, respectively.  This might open the possibility to simulate
the nuclear tensor force in experiments with ultracold atoms.

%% file: chpt.tex
\section{Three-body forces in few-nucleon systems}
\label{sec:chpt}

\begin{figure}[t]
\begin{center}
\includegraphics[width=3cm,clip=]{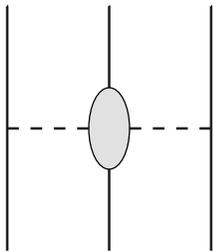}
\end{center}
\caption{Topology of the two-pion-exchange 3NF. Solid (dashed) lines indicate
nucleons (pions).}
\label{tnf2pi}
\end{figure}

Three-body forces are especially important in nuclear
physics. Phenomenological studies indicate, e.g., that the
contribution of 3NFs to binding energies of light nuclei is
quantitatively significant, of the order 20\% (see
\cite{Pieper:2001mp}).

In nuclear systems, the long-range parts of nuclear forces are
mediated by pion exchanges in addition to short-range contact
interactions discussed in the preceding section. The pion mass is
comparable to the momenta in typical nuclei. Therefore, it cannot be
expected that pionless EFT is applicable, unless one considers very
specific observables, e.g., for halo nuclei where the nucleon
separation energy presents a low scale.

An important signature of 3NFs is the model dependence of few-nucleon
predictions when only NN interactions are employed in the calculation.
This is demonstrated by the $^3$H and $^4$He binding energies in
Fig.~\ref{fig:tjon}. The model dependence indicates that the missing
three-nucleon and higher-body interactions are different for each NN
potential employed
(\cite{Polyzou:1990isi,Amghar:1995av}). Fortunately, 4N and
higher-body forces are expected to be further suppressed.  We will
come back to this issue, but assume for the moment that such
contributions are negligible.

It is therefore required to formulate both two- and three-body forces
within one systematic scheme. Historically, this has not been the
case. In most models, the main contribution is related to the
two-pion-exchange contribution depicted in Fig.~\ref{tnf2pi}.
\cite{Fujita:1957zz} realized that this model can be constrained using
pion-nucleon scattering data and found that the interaction is
dominated by P-wave pion-nucleon ($\pi$N) interactions. This was the
birth of modern 3NFs which were mostly developed independently of NN
interactions (see, e.g., \cite{Coon:1978gr}). Two-pion-exchange 3NFs
are generally of the form (see \cite{Friar:1998zt}):
\begin{equation}
\label{eq:2pi}
V^{\rm 3NF}_{2\pi} = \frac{1}{2} \sum_{i \neq j \neq k}
\frac{g_{A}^2}{(2 f_{\pi})^{2}}
{\vec \sigma_{i} \cdot \vec q_{i} \, \vec \sigma_{j} \cdot \vec q_{j} \over 
( {\vec q_{i}} \, ^{2} +m_{\pi}^{2} ) ( {\vec q_{j}} \, ^{2} +m_{\pi}^{2} ) } 
F_{ijk}^{\alpha \beta} \tau_{i}^{\alpha} \tau_{j}^{\beta} ,
\end{equation}
with
\beqa
F_{ijk}^{\alpha \beta} &=& \delta_{\alpha \beta} 
\left[ -{4 c_{1} m_{\pi}^{2} \over f_{\pi}^{2}} 
+ {2 c_{3}  \over f_{\pi}^{2}} \  \vec q_{i} \cdot \vec q_{j}  \right]
\nonumber \\[5pt]
&&+ \, {c_{4}  \over f_{\pi}^{2}} \ \epsilon^{\alpha \beta \gamma} \,
\tau_{k}^{\gamma} \, \vec \sigma_{k}
\cdot ( \vec q_{i} \times \vec q_{j}) \,,
\eeqa
where $i,j,k$ label particles and $\alpha, \beta, \gamma$ isospin,
$m_{\pi}$ is the pion mass, $f_{\pi}=92.4 \, {\rm MeV}$ the pion decay
constant, and $g_{A}$ the axial pion-nucleon coupling.  The constants
$c_{i}$ are different for all models. In these equations, we have
neglected cutoff functions that are required to regularize 3NFs at
short distances.

In these models, the 3NFs are unrelated to the NN interaction, which
shows up in a strong model dependence of predictions based on
combining such 3NFs and different NN interactions. Although often
parts of the parameters are adjusted using $\pi$N scattering data, one
still needs to adjust a parameter of the 3NF, e.g., a cutoff
parameter, such that the prediction for the $^{3}$H binding energy
agrees with experiment. Such combinations are not based on a
consistent framework. They do not describe all available 3N scattering
data, but they improve the description of many low-energy few-nucleon
observables (\cite{KalantarNayestanaki:2011wz}). In part, this is
related to the universal correlations of observables discussed in the
last section. Such observables are thus not useful to pin down the
spin-isospin structure of 3NFs.

Therefore, investigations have concentrated on intermediate-energy
nucleon-deuteron scattering. Using phenomenological forces, it can be
shown that, for this energy range, observables exist that are
sensitive to the structure of 3NFs (\cite{Witala:2001by}).  Due to a
series of sophisticated nucleon-deuteron scattering experiments, data
is now available (\cite{KalantarNayestanaki:2011wz}).  Unfortunately,
this data has not been analyzed yet in a framework that provides
consistent NN and 3N interactions. Comparison of the data to the
predictions based on phenomenological forces show that the current
models do not describe the intermediate-energy data very well.  We
note that improvements of the models have been suggested (see, e.g.,
\cite{KalantarNayestanaki:2011wz} for more details).  Here, we will
focus on selected low-energy observables within a systematic approach
to nuclear forces.

\begin{figure}[t]
\begin{center}
\includegraphics[width=5cm,clip=]{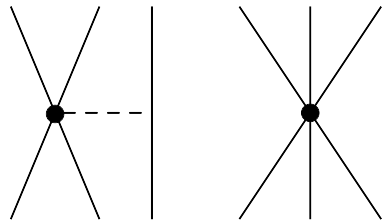}
\end{center}
\caption{Topology of the leading mid-range (left) and short-range
(right) 3NFs. Solid (dashed) lines indicate
nucleons (pions).}
\label{tnfshort}
\end{figure}

Such a systematic approach has been developed based on chiral EFT (for
recent reviews, see \cite{Epelbaum:2005pn,%
Epelbaum:2008ga,Machleidt:2011zz}).  Based on the symmetries of the
QCD Lagrangian, most importantly its spontaneously broken chiral
symmetry, it is possible to formulate an EFT in terms of nucleon
degrees of freedom and the nearly massless Goldstone bosons of QCD:
the pions. Symmetry considerations sufficiently constrain the
interactions of pions with themselves and with nucleons to develop a
systematic power counting scheme: chiral perturbation theory (ChPT).
The expansion parameter is $Q/\Lambda_{\chi}$, where $Q \sim m_\pi$ is
a typical momentum or the pion mass and $\Lambda_{\chi} \approx 1 \,
{\rm GeV}$ is the chiral-symmetry breaking scale. The terms of the
Lagrangian that are relevant to our discussion here read
\begin{eqnarray}
\label{eq:lagrange}
{\cal L} &=& \psi^\dagger \left[ 4 c_{1} m_{\pi}^{2} 
- {2 c_{1} \over f_{\pi}^{2}} \, m_{\pi}^{2} \vec \pi \, ^{2}
+ { c_{3} \over  f_{\pi}^{2} } \, \partial_{\mu} \vec \pi \, 
\cdot \partial^{\mu} \vec \pi \,  \right. 
\nonumber \\[5pt]
&&\quad\left. - { c_{4} \over 2 f_{\pi} ^{2}}  
\, \epsilon_{ijk} \, \epsilon_{abc} \, \sigma_{i} \, \tau_{a} \,
(\nabla _{j}  \pi_{b} ) (\nabla _{k}  \pi_{c} ) \right] \psi
\nonumber \\[5pt]  
&&- \, {D \over 4 f_{\pi}} (\psi^{\dagger} \psi) ( \psi^{\dagger} 
\vec \sigma \vec \tau  \psi) \cdot  \vec \nabla \pi
\nonumber \\[5pt]
&&- \, {E \over 2} (\psi^{\dagger} \psi) (\psi^{\dagger} \vec \tau
\psi) (\psi^{\dagger} \vec \tau  \psi) + \ldots \,,
\end{eqnarray}
where $\psi$ and $\vec \pi$ are nucleon and pion fields, respectively.
The Lagrangian includes as $\pi\pi$NN vertices the same LECs $c_{i}$
of the two-pion-exchange 3NFs. In addition, these LECs also contribute
to the subleading two-pion-exchange NN interaction, which shows the
strong connection of NN and 3N forces.

The challenge for nuclear forces is that, because of bound state,
parts of the interactions are nonperturbative in contrast to the
interactions in pionic or $\pi$N systems. This issue has been tackled
assuming that the power counting can be applied to all non-reducible
diagrams without purely nucleonic intermediate states. These diagrams
form a potential that is then summed to all orders solving the
Schr\"odinger equation (\cite{Weinberg:1990rz}). The approach requires
numerical regularization of the potential introducing a regulator
dependence.  In state-of-the-art applications, cutoffs are presently
restricted to values $\Lambda \lesssim 450 - 600 \, {\rm MeV}$. There
is a lively discussion whether this constraint is an artifact and can
be overcome by an improved power counting, or whether it is an
inherent requirement of a nonperturbative extension of ChPT
(\cite{Nogga:2005hy,Hammer:2006qj,Epelbaum:2009sd,Valderrama:2009ei}).
For applications to many-nucleon systems, lower cutoffs are
advantageous and in some cases, computations are only feasible with
lower cutoffs. For the estimate of missing higher-order contributions,
we will assume a high scale $\Lambda_\chi \approx \Lambda$ in this
section.

In chiral EFT, NN, 3N and higher-body forces can be derived
consistently. The general result is that higher-body forces are
suppressed compared to lower-body ones. This justifies our assumption
that 4N and higher-body forces are further suppressed. For NN
interactions, one finds that the longest-range part is one-pion
exchange, which is also the basis of state-of-the-art NN models. The
first 3NF contribution is suppressed by $(Q/\Lambda_{\chi})^3$
(\cite{vanKolck:1994yi}) and contains the two-pion-exchange part given
by Eq.~(\ref{eq:2pi}). At the same order of the expansion, two other
topologies (see Fig.~\ref{tnfshort}) contribute
\begin{equation}
\label{eq:short}
V^{\rm 3NF}_{\rm short} = \sum_{i \neq j \neq k} 
\left[ - D \ {g_{A} \over 8 f_{\pi}^{2} } 
\, {\vec \sigma_{j} \cdot \vec q_{j} \, \vec \sigma_{i} \cdot \vec
q_{j} \over \vec q_{j} \, ^{2} + m_{\pi}^{2}} \, 
\vec \tau_{i} \cdot \vec \tau_{j} 
+ {E \over 2} \, \vec \tau_{j} \cdot \vec \tau_{k} \right] .
\end{equation} 
Usually, these parts are called the $D$- and $E$-term. The $D$-term is
of mid range (one-pion-exchange--short-range) and the $E$-term is of
short range. This implies that the $E$-term coupling can only be
obtained from few-nucleon observables, whereas the $D$-term strength
is also related to weak or pionic processes involving two nucleons
(see Section~\ref{sec:other}).  Following standard conventions, we
introduce two dimensionless couplings $c_D= D \ f_{\pi}^{2}
\Lambda_{\chi}$ and $c_E= E \ f_{\pi}^{4} \Lambda_{\chi}$.  As noted
above, due to the $c_{i}$ vertices of the Lagrangian, ChPT provides
relations between the strength of the two-pion-exchange NN interaction
and $V^{\rm 3NF}_{2\pi}$. This level of consistency can only be
implemented in the framework of ChPT.  For the results given here, the
$\Delta$ is not treated as an explicit degree of freedom. Since the
mass difference of the nucleon and the $\Delta$ is only $\sim 2
m_\pi$, an explicit inclusion is expected to improve the convergence
of the chiral expansion
(\cite{Ordonez:1993tn,Kaiser:1998wa,Krebs:2007rh}).  For 3NFs, the
leading $\Delta$ contribution is entirely included in $V^{\rm
3NF}_{2\pi}$ and shows up in larger strength constants $c_{3}$ and
$c_{4}$ enhancing the two-pion-exchange contributions compared to the
other two topologies (\cite{Epelbaum:2007sq}).

In nuclear systems the separation of the high and the low scales
(given by $\Lambda$ and the pion mass or a typical momentum of the
system, respectively) is not exceedingly large, which implies a slowly
converging chiral expansion. Especially for intermediate-energy
nucleon-deuteron scattering, the expansion parameter is estimated to
be $\sim 1/2$ or larger. Therefore, calculations up to order $Q^3$
(including the leading 3NFs) are useful only up to nucleon laboratory
energies of $\sim 100 \, {\rm MeV}$.  Fortunately, the $Q^4$ 3NF
contributions have been completed recently
(\cite{Bernard:2007sp,Bernard:2011zr}), and applications are under way.

\begin{table}[t]
\caption{Comparison of different $c_{i}$ determinations. The $c_{i}$'s
are given in GeV$^{-1}$. At present, the determinations using NN observables
require further constraints from $\pi$N observables to be conclusive.
The last column indicates whether the $c_{i}$ values are mostly based
on NN or $\pi$N data. We also show the results based on resonance
saturation (res) from \cite{Bernard:1996gq} (note that we omitted the 
$\pi$N fit from that paper).\label{tab:cis}}
\begin{center}
\setlength{\tabcolsep}{5pt} 
\begin{tabular}{l|c|c|c|c}
& $c_{1}$ & $c_{3}$ & $c_{4}$ & \cr \hline
\cite{Fettes:1998ud} (Fit 1) & -1.2 & -5.9 & 3.5 & $\pi$N \cr 
\cite{Buettiker:1999ap} & -0.8 & -4.7 & 3.4 & $\pi$N \cr 
\cite{Meissner:2007trf} & -0.9 & -4.7 & 3.5 & $\pi$N \cr 
\cite{Rentmeester:2003mf} & -0.8 & -4.8 & 4.0 & NN \cr 
\cite{Entem:2002sf} & -0.8 & -3.4 & 3.4 & NN \cr 
\cite{Entem:2003ft} & -0.8 & -3.2 & 5.4 & NN \cr
\cite{Epelbaum:2004fk} & -0.8 & -3.4 & 3.4 & NN \cr 
\cite{Bernard:1996gq} & -0.9 & -5.3 & 3.7& res
\end{tabular}
\end{center}
\end{table}

Before calculations based on chiral 3NFs can be performed, one needs
to determine the LECs $c_{i}$, $c_{D}$ and $c_{E}$.  The $c_{i}$ constants
have been determined from NN data as well as $\pi$N data. The results
are summarized in Table~\ref{tab:cis}. For simplicity, we have omitted
the theoretical uncertainties and only give the central values. Most
determinations are in agreement within the uncertainties, but
deviations of the different determinations can be sizable, of the order
of 30\%. For our purpose here, this accuracy is sufficient and
comparable to higher-order contributions that we do not take into
account. This problem will become more relevant, when
the subleading parts of the 3NF will allow us to increase the accuracy of
our predictions. In principle, these constants can be
obtained independently of the NN interaction. So their size should not
depend on the regulator chosen or on the specific realization of
chiral NN potentials. 

Based on naturalness arguments, one would expect that the $c_{i}$'s
are of the order of $\Lambda_\chi^{-1} \sim 1 \, {\rm Gev}^{-1}$.  It
sticks out that $c_{3}$ and $c_{4}$ are larger than this estimate.
This can be understood based on resonance saturation, where the large
$c_{i}$'s are related to the small $\Delta$ to nucleon mass difference
$\sim 1/(m_\Delta - m)$ (see, e.g., \cite{Bernard:1996gq}).  Taking
$\Delta$'s explicitly into account reduces the magnitude of the
$c_{i}$ considerably so that an improved convergence of the chiral
expansion can then be expected (see Section~\ref{outlook} and
\cite{Krebs:2007rh} in the context of NN interactions).

Finally, we need to determine the constants $c_{D}$ and
$c_{E}$. Usually, combinations of $c_{D}$ and $c_{E}$ are found that
make sure that the $^3$H binding energy is described correctly. Then
different strategies have been used to constrain $c_{D}$ from
few-nucleon data, e.g., by fitting the doublet neutron-deuteron
scattering length (\cite{Epelbaum:2002vt}), the binding energy of
$^4$He (\cite{Nogga:2005hp}), or the radius of $^4$He
(\cite{Navratil:2007we}). In addition, the $^3$H beta decay half-life
can be used to constrain $c_D$ (see Section~\ref{sec:other}). In
particular, the fit of $c_D$ to the $^3$H beta decay half-life or to
the radius of $^4$He have been shown to lead to a good overall
description of light and $p$-shell nuclei. It is important to note
that many low-energy observables are already well described once
$c_{D}$ and $c_{E}$ combinations have been chosen that describe the
$^3$H binding energy correctly. Therefore, the sensitivity of
these observables on $c_{D}$ is low and a considerable uncertainty
remains. Possibly, for higher-order calculations, other strategies
need to be devised to obtain more accurate determinations of $c_{D}$
and $c_{E}$.

\begin{table}[t]
\caption{Power counting predictions and explicit results for the
binding energy $B$ and the expectation values of NN and 3N forces
for $^4$He. Cutoffs and energies are given in MeV.\label{tab:exp3nf}} 
\begin{center}
\setlength{\tabcolsep}{5pt} 
\begin{tabular}{c|c|c|c|c}
$\Lambda$ & $B$ & $\langle V_{\rm NN} \rangle$ 
& $\langle V_{\rm 3NF} \rangle$ & ${ \left| \langle V_{\rm 3NF} \rangle
\over \langle V_{\rm NN} \rangle \right| }$ [\%] \cr \hline
450 & 27.65 & $-84.56$ & $-1.11$ & 1.3 \cr 
600 & 28.57 & $-93.73$ & $-6.83$ & 7.2
\end{tabular}
\end{center}
\end{table}

Since the separation of scales is not very large and since there are
ongoing discussions on the size of the high scale for nuclei, it is
instructive to calculate the contributions of NN and 3N forces to the
binding energy of light nuclei. These contributions are not
observables, nevertheless their relative size can be estimated and
compared to the power counting estimate. For this estimate, we use the
realization of chiral EFT interactions at order $Q^3$ of
\cite{Epelbaum:2004fk} (with $\tilde \Lambda = 700$~MeV).  In this
work, the NN potential has been fitted for different cutoffs, which
can be used to investigate the scale dependence of chiral 3NFs.  For
the 3NF, we use the same $c_{i}$ values as for the NN part.  The
$c_{D}$ and $c_{E}$ values have been determined by a fit to the $^3$H
binding energy and the doublet neutron-deuteron scattering length. The
chiral 3NFs have been regularized using a cutoff function depending on
the relative momenta in the in- and outgoing state where the cutoff is
identical to the $\Lambda$ of \cite{Epelbaum:2004fk}. The results are
given in Table~\ref{tab:exp3nf}. As one can see, the binding energy of
$^4$He is close to the experimental value of $28.30 \, {\rm MeV}$. The
remaining deviation from experiment is comparable to the cutoff
dependence and indicates the contribution that can be expected from
order $Q^4$. The leading 3NF is a $Q^{3}$ contribution. Assuming a
typical momentum $\sim m_{\pi}$ and $\Lambda_{\chi} = 500 \, {\rm
MeV}$, we expect a contribution of approximately 2\% to the
potential energy.  It is apparent that the contribution of 3NFs
strongly depends on the cutoff. For the first case in
Table~\ref{tab:exp3nf}, the size is smaller than expected, which is no
contradiction to the power counting. For the second case, the 3NF
contribution is somewhat larger than naively expected. The estimate is
still within a factor of $3-4$ correct (a natural-sized number), but it
shows the enhancement of the 3NF due to the $\Delta$ resonance.

In summary, the overall size of 3NF contributions is as
expected from the power counting once the contribution of the $\Delta$
resonance has been taken into account. The deviation of the binding
energy for $^4$He can also be expected from a higher-order
contribution. On a quantitative level, this deviation indicates that
high precision can only be expected for a $Q^4$ calculation.

\begin{figure*}[t]
\begin{center}
\includegraphics[height=5.0cm,clip=]{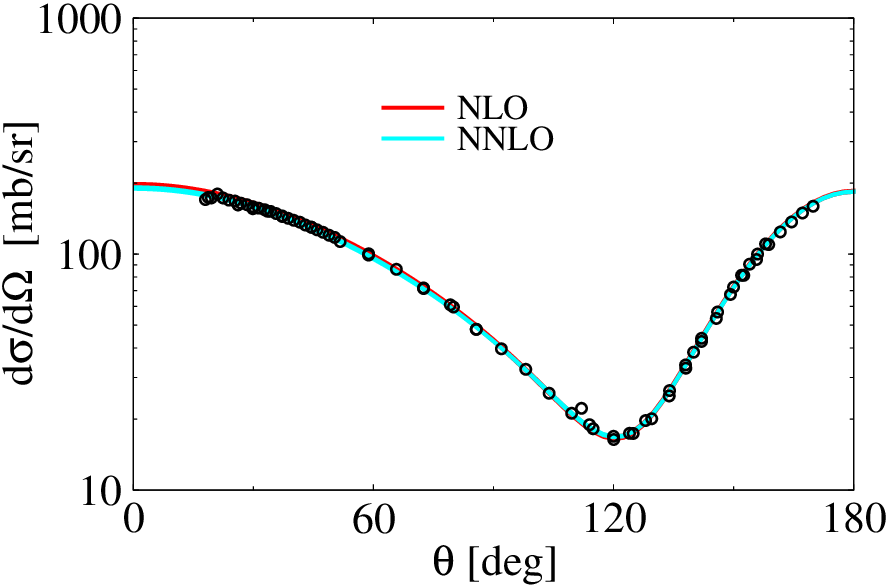}
\hspace{0.5cm}
\includegraphics[height=5.0cm,clip=]{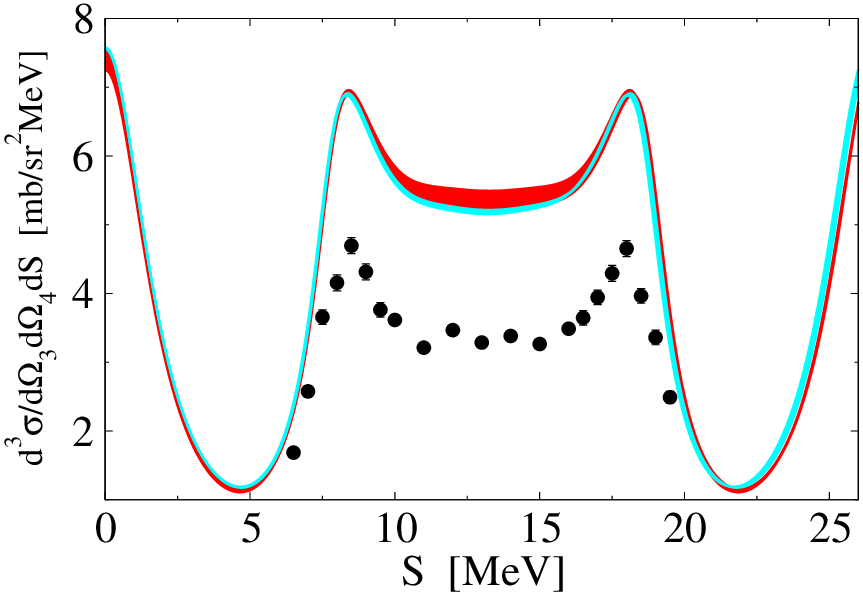}
\end{center}
\caption{(Color online) Left panel: Elastic nucleon-deuteron cross 
section at $10 \, {\rm MeV}$. The almost indistinguishable bands 
correspond to chiral $Q^{2}$ (red, dark grey) and $Q^{3}$ (cyan, 
light grey) calculations. Data are from
\cite{Howell:1987fbs,Sagara:1994zz,Rauprich:1988fbs,Sperisen:1984npa}.
Right panel: Nucleon-deuteron breakup cross section at $19 \, 
{\rm MeV}$ for the space-star configuration at $\alpha=56^\circ$ 
(see \cite{Ley:2006hu} for the definition of the kinematics).
The bands are the same as in the left panel. The data is 
from \cite{Ley:2006hu}.\label{fig:ndxsec}}
\end{figure*}

At order $Q^3$, there are also nucleon-deuteron scattering
calculations available. At intermediate energies, the results are
strongly dependent on the cutoff. For low energy, however, many
observables can be accurately predicted. In the left panel of
Fig.~\ref{fig:ndxsec}, we show as an example the elastic
nucleon-deuteron cross section. For the elastic cross section, data
and prediction are in excellent agreement, and the order $Q^2$ and
$Q^3$ results are similar indicating that the calculation is converged
with respect to the chiral expansion. Whereas the bulk of the
observables at low energies are nicely reproduced, there are a some
exceptions. One of them is a specific breakup configuration shown in
the right panel of Fig.~\ref{fig:ndxsec}. Again, the $Q^2$ and $Q^3$
results nicely agree indicating convergence of the chiral expansion
and, therefore, small contributions of 3NFs. Unfortunately, there is a
large discrepancy to the data. This is still an unresolved problem.

\begin{figure}[t]
\begin{center}
\includegraphics[height=5.75cm,clip=]{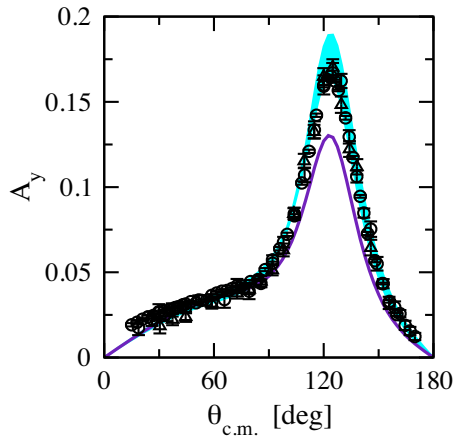} 
\end{center}
\caption{(Color online)
The nucleon vector analyzing power $A_{y}$ for elastic
nucleon-deuteron scattering at $10 \, {\rm MeV}$. The cyan band covers
predictions based on chiral NN and 3N forces at order $Q^3$ for
various cutoffs. The solid, purple line is the result of the CD-Bonn
NN combined with the TM99 3N potential. Data as in the left panel
of Fig.~\ref{fig:ndxsec}. Reprinted with permission from 
\cite{KalantarNayestanaki:2011wz}.\label{fig:ay}}
\end{figure}

A different example is the analyzing power at low energy shown in
Fig.~\ref{fig:ay}. The solid line represents the result based on
high-precision phenomenological forces, which clearly disagrees with
the data.  This is a common feature of all available calculations
based on phenomenological forces and has been discussed vividly in the
literature (see, e.g., \cite{Miller:2007ad}).  At order $Q^3$, for the
realization of \cite{Epelbaum:2004fk}, the result seems to agree with
the data.  However, the cutoff dependence is unusually large for these
small energies and a detailed analysis reveals that this agreement in
the three-body sector can be traced back to deficiencies in the
description of NN data.  Therefore, at order $Q^3$, the analyzing
power cannot be properly predicted but is merely accidentally
described.  We stress that the analyzing power is a very small
observable, so that tiny improvements of the Hamiltonian can be
relevant for a proper prediction. This is in line with the rather
strong dependence on the cutoff, which indicates that order $Q^4$
contributions might resolve this puzzle. For the 4N system, a more
significant deviation of data and predictions of the analyzing power
has been found by \cite{Viviani:2001cu}.  Interestingly, in this case,
chiral 3NFs lead to an improved description of the data compared to
the standard phenomenological forces (\cite{Viviani:2010mf}).

In summary, the results for few-nucleon systems show that $Q^3$
predictions are in line with the expectations based on power
counting. Whereas low-energy scattering is reasonably described at
this order, the results for the binding energies indicate that $Q^4$
will be required to reach satisfactory accuracies. Two-nucleon forces
at this order are available and have an accuracy comparable to
phenomenological forces (\cite{Entem:2003ft,Epelbaum:2004fk}). The
3NFs at N$^3$LO have been completed recently
(\cite{Bernard:2007sp,Bernard:2011zr}). In addition, a consistent
calculation up to this order also involves 4N forces, which
fortunately do not involve additional LECs and are therefore parameter
free. They have been derived and explored in $^4$He
(\cite{Epelbaum:2007us,Nogga:2010epj}).  In this case, for the small
cutoffs, the contributions seems to be smaller than expected. It
remains to be seen whether this is also true for more complex systems.

Next, we return to the discussion of the RG transformation started in
Section~\ref{sec:def}, because the resolution scale dependence also
applies to the low-energy couplings in 3NFs. Therefore, the RG
equation in the NN sector, Eq.~(\ref{eq:nnrg}) needs to be augmented
by a similar equation for 3NFs, which we again write schematically as
\begin{equation}
\label{eq:3nrg}
\frac{d}{d \Lambda} \, V_{\Lambda}(123) = F_{123}(V_{\Lambda}(ij),
V_{\Lambda}(123),\Lambda) \,.
\end{equation}
For low-momentum interactions $\Vlowk$, solving the RG equation for
3NFs is difficult in practice, because it involves a complete set of
scattering solutions for the three-body system (\cite{Bogner:2009bt}).
This is not feasible at this point, but a consistent 3NF evolution can
be carried out in the SRG approach (as discussed below). For
low-momentum interactions, the chiral EFT has been used as a general
operator basis of 3NFs, and the LECs have been adjusted directly to
few-nucleon data at lower resolution scales (\cite{Nogga:2004ab,%
Hebeler:2010xb}). Such an approach is justified for $\Lambda
\lesssim 500 \, {\rm MeV}$, because the NN interactions become
universal (\cite{Bogner:2003wn}). It is therefore motivated that
consistent 3NFs should have the same form as the ones derived in
chiral EFT. Since 3NFs are defined up to a finite order, even
three-nucleon observables will only be approximately independent of
the cutoff. It is therefore common to perform calculations for a range
of cutoff values. The variation of the prediction provides an estimate
of the theoretical uncertainty due to neglected higher-order 3NFs.  If
observables are calculated for more complex systems, the dependence on
the cutoff can also be due to neglected four- and higher-body
interactions.

The SRG approach provides a powerful scheme to evolve 3NFs by
differential equations of the general form of Eqs.~(\ref{eq:nnrg})
and~(\ref{eq:3nrg}).  The SRG transformation is an exact unitary
transformation. Therefore, all NN observables are invariant under the
transformation. By construction, low and high momenta decouple, and
observables at low momentum becomes insensitive to high-momentum
details (\cite{Jurgenson:2007td}). As a result, many-body calculations
converge more rapidly for evolved potentials, similar to low-momentum
interactions.  The SRG evolution of 3NFs has been achieved in a
harmonic-oscillator basis (\cite{Jurgenson:2009qs,Roth:2011ar}) and
recently in momentum space (\cite{Hebeler:2012pr}).

\begin{figure}[t]
\begin{center}
\includegraphics[width=7cm,clip=]{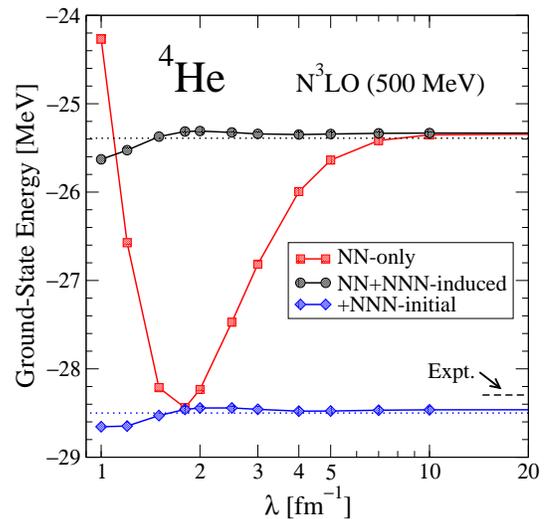}
\end{center}
\caption{(Color online)
Ground-state energy of $^{4}$He as a function of the SRG flow
parameter $\lambda$ starting from chiral NN and 3N interactions at
N$^3$LO and N$^2$LO, respectively. For details see \cite{Jurgenson:2009qs}.
\label{fig:4hesrg}}
\end{figure}

The results of the application of SRG-evolved NN and 3N interactions
to $^{4}$He again shows the quantitative importance of 3NFs for
binding energies of nuclei. But it also supports the general belief
that four-nucleon forces do not contribute significantly, as shown in
Fig.~\ref{fig:4hesrg}.  When the SRG flow is truncated at the two-body
level, the ground-state energy of $^{4}$He depends significantly on
the SRG flow parameter $\lambda$, which plays a similar role as the
momentum cutoff $\Lambda$. However, the $\lambda$ variation is of the
same order as the 3NF contribution. This shows how the RG/SRG cutoff
variation estimates missing parts of the Hamiltonian. When 3NFs are
included in the SRG evolution, most of the $\lambda$ dependence is
removed. The remaining variation is of the order of $50 \, {\rm keV}$
for $\lambda \gtrsim 1.5 $~fm$^{-1}$, indicating that induced 4N
forces provide a small contribution to the $^{4}$He ground-state
energy. Note that this estimate is even smaller than explicit
calculations using 4N forces (\cite{Deltuva:2008jr,Nogga:2010epj}),
which result in $200-300$~keV for $^4$He.

The small size of 4N forces justifies the exploration of larger nuclei
and nuclear matter based on chiral NN and 3N interactions in the next
Section. The detailed calculation of 4N forces indicates that the
result for $^4$He might be suppressed because parts of the force
cancel for these quantum numbers. Eventually, this result needs to be
confirmed for more complex systems than $^4$He.

%% file: manybody.tex
\section{Three-nucleon forces and many-body systems}
\label{sec:manybody}

Three-body forces are a frontier for understanding and predicting
strongly interacting many-body systems. While the quantitative
importance of 3NFs has been well established in light nuclei, they are
currently not included in most nuclear structure calculations. In this
section, we discuss the opportunities and challenges this area offers.
We highlight the importance of 3NFs beyond light nuclei, for
neutron-rich systems, and for nucleonic matter in astrophysics, with a
focus on 3NFs based on chiral EFT. Although some of the applications
that we discuss still require an approximative treatment of 3NFs, they
exhibit new facets and significant contributions of 3NFs.

As discussed in the previous section, chiral EFT opens up a systematic
path to investigate many-body forces, which has not been possible
before. This results from the consistency of NN and 3N interactions
and the possibility to constrain all parameters using only few-nucleon
data. No new parameters enter for 3N and 4N forces at N$^3$LO.
Moreover, it has been shown that for systems of only neutrons, the $D$
and $E$ parts do not contribute because of the Pauli principle and the
coupling of pions to spin (\cite{Hebeler:2009iv,Tolos:2007bh}).  This
establishes a forefront connection of the investigation of 3NFs with
the exploration of neutron-rich nuclei at rare isotope beam facilities
worldwide.

\begin{figure}[t]
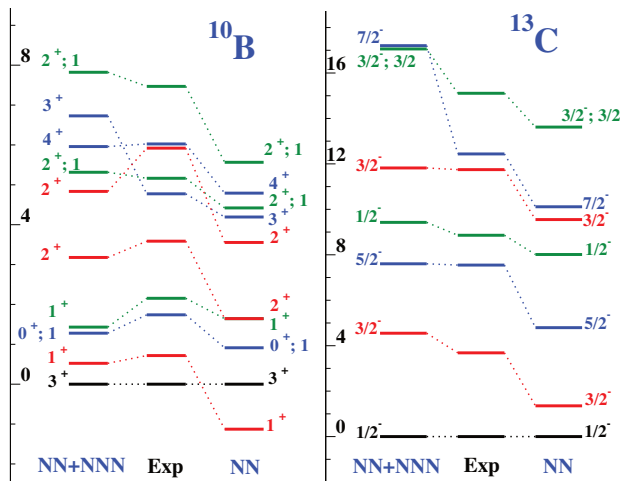

\begin{center}
\includegraphics[width=0.225\textwidth,clip=]{B10.eps}
\includegraphics[width=0.225\textwidth,clip=]{C13.eps}
\end{center}
\caption{(Color online) Excitation energies in MeV of light nuclei,
$^{10}$B and $^{13}$C, obtained in the ab-initio No-Core Shell Model
(NCSM) with chiral EFT interactions (NN to N$^3$LO and 3N to N$^2$LO) 
(\cite{Navratil:2007we}).\label{pshell}}
\end{figure}

\begin{figure*}[t]
\begin{center}
\includegraphics[width=0.98\textwidth,clip=]{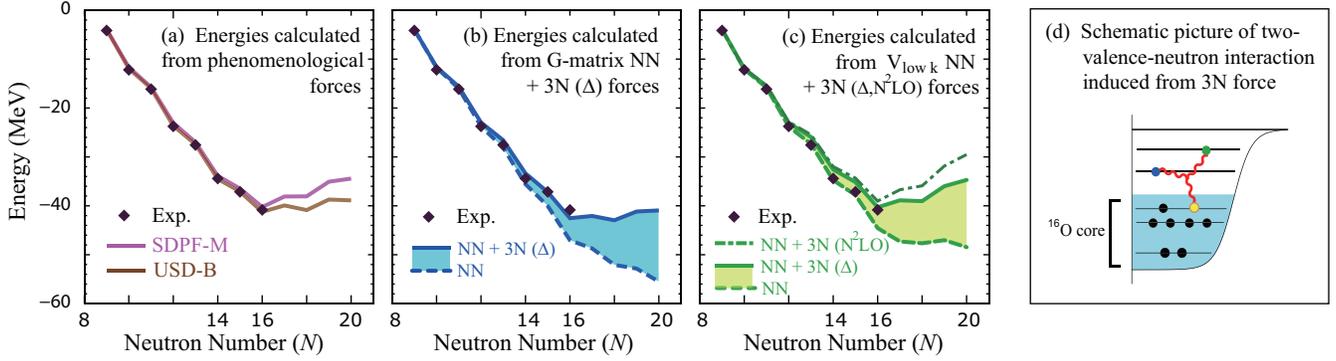}
\end{center}
\caption{(Color online) Ground-state energies of the neutron-rich
oxygen isotopes relative to $^{16}$O, including experimental values
of the bound isotopes $^{16-24}$O. Energies obtained from
(a)~phenomenological forces SDPF-M and USD-B, (b)~a $G$ matrix
interaction and including Fujita-Miyazawa 3NFs due to $\Delta$
excitations, and (c)~from low-momentum interactions $\Vlowk$ and
including chiral 3NFs at N$^2$LO as well as only due to $\Delta$
excitations. The changes due to 3NFs based on $\Delta$ excitations
are highlighted by the shaded areas. (d)~Schematic illustration of a
two-valence-neutron interaction generated by 3NFs with a nucleon in
the $^{16}$O core. For details see \cite{Otsuka:2009cs}.\label{oxygen}}
\end{figure*}

As expected from the Tjon band in Fig.~\ref{fig:tjon}, 3NFs impact
binding energies and therefore also radii. Precision techniques for
masses and charge radii present new challenges for theory (in the
context of 3NFs see, e.g., \cite{Brodeur:2011kn}). In addition,
similar to the spin dependences observed in few-body scattering, e.g.,
for the analyzing power $A_y$ discussed in Fig.~\ref{fig:ay}, 3NFs
play an important role for spin-orbit splittings and spin dependences
in nuclei. Both aspects can be clearly seen in the spectra of light
nuclei, where calculations can be performed ab-initio, making these
nuclei an interesting laboratory to explore nuclear forces. As an
example, we show two representative spectra in Fig.~\ref{pshell},
calculated in the No-Core Shell Model (NCSM) including chiral 3NFs at
N$^2$LO (\cite{Navratil:2007we}). The NCSM is based on a large-basis
Hamiltonian diagonalization. Without 3NFs the spectra are generally
too compressed (which is also found for $^{23}$O in Fig.~\ref{oxygen}
and for other medium-mass nuclei). Clearly, the spectrum improves,
when 3NFs are taken into account.  In addition to a repulsive effect
on the spectra, 3NFs provide important contributions to the spin-orbit
splitting, reflected in the excitation energy of the first
$\nicefrac{3}{2}^-$ state relative to the $\nicefrac{1}{2}^-$ ground
state in $^{13}$C, which probes the splitting of the
$p_{\nicefrac{3}{2}}-p_{\nicefrac{1}{2}}$ orbitals. This can also be
seen in the 3NF contributions to the spin-orbit splitting between the
$p_{\nicefrac{1}{2}}$ and $p_{\nicefrac{3}{2}}$ phase shifts in
nucleon-$^4$He scattering (\cite{Nollett:2006su}).  For $^{10}$B, NN
forces alone do not predict the correct ground-state spin and parity
$3^+$, but instead the lowest state is found to be $1^+$.  This is
only corrected by some of the phenomenological 3NFs
(\cite{Pieper:2007ax}). For chiral 3NFs, the correct ordering is
predicted. This is also needed for the analogous states in medium-mass
nuclei $^{22}$Na and $^{46}$V, which are $N=Z=8$ and $N=Z=20$ nuclei
with three valence neutrons and three valence protons
(\cite{Nowacki:2008tlk}). Moreover, recent work has demonstrated the
impact of 3NFs on the structure probed in electroweak transitions (see
also Section~\ref{sec:other}), e.g., for the beta decay of $^{14}$C
used for carbon dating (\cite{Holt:2009uk,Maris:2011as}).

Nuclear lattice simulations were recently used to perform the first
ab-initio calculation of the Hoyle state in $^{12}$C
(\cite{Epelbaum:2011md}), which is important for nucleosynthesis.  Due
to its alpha-cluster structure, this state is challenging for
many-body methods.  In this approach, spacetime is discretized and the
nucleons are located on the lattice sites. Their interactions in
chiral EFT are implemented using auxiliary fields and the low-lying
states are extracted using a generalized Euclidean time projection
method. This promising new method allows to take 3NFs into account
without handling large interaction matrices.

The application of RG transformations to evolve nuclear forces to
lower resolution leads to greatly enhanced convergence in few- and
many-body systems (\cite{Bogner:2009bt}). Current research focuses on
extending these methods to 3NFs using the SRG. This has been achieved
in a harmonic-oscillator basis (\cite{Jurgenson:2009qs}; see
Fig.~\ref{fig:4hesrg}), with very promising results in light and
medium-mass nuclei in the NCSM and importance-truncated NCSM
(\cite{Jurgenson:2010wy,Roth:2011ar}), and recently in momentum space
(\cite{Hebeler:2012pr}). Open questions include understanding the
cutoff dependence in chiral EFT, whether long-range many-body
interactions are induced by the SRG, and to explore the dependence on
the SRG generator.

\begin{figure}[t]
\begin{center}
\includegraphics[scale=0.35,clip=]{23O_rev2.eps}
\end{center}
\caption{Excitation energies of bound excited states in $^{23}$O
compared with experiment (\cite{Elekes:2007zz,Schiller:2006hz}). The
NN-only results are calculated in the $sd$ and
$sdf_{\nicefrac{7}{2}}p_{\nicefrac{3}{2}}$ shells with empirical
single-particle energies (SPEs). The NN+3N energies are obtained in
the same spaces, but with calculated SPEs including 3NFs at N$^2$LO.
For details see \cite{Holt:2011fj}. The dashed lines
give the one-neutron separation energy $S_n$.\label{23O}}
\end{figure}

\begin{figure*}[t]
\begin{center}
\includegraphics[scale=0.45,clip=]{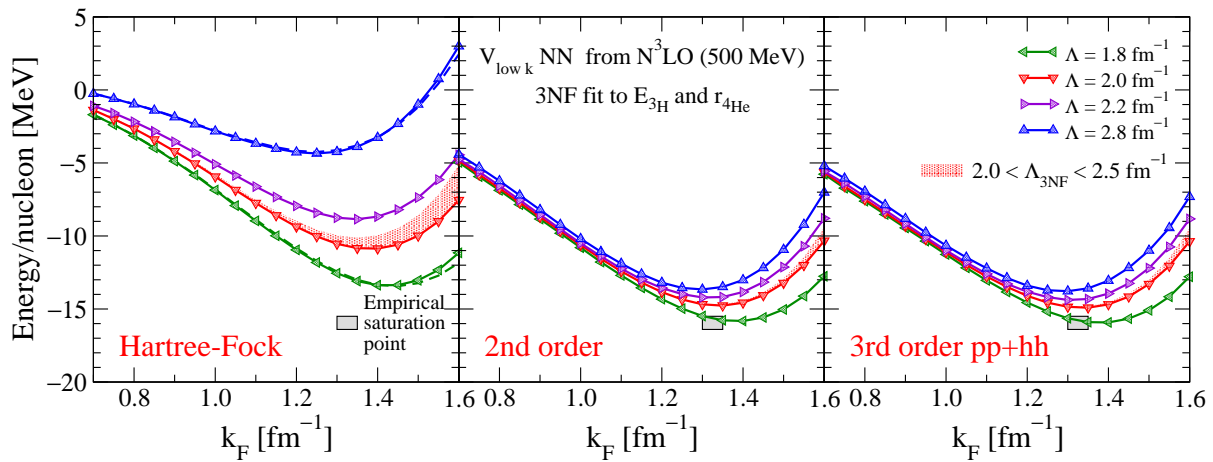}
\end{center}
\caption{(Color online) Nuclear matter energy per particle versus
Fermi momentum $k_{\rm F}$ at the Hartree-Fock level (left) and
including second-order (middle) and third-order
particle-particle/hole-hole contributions (right), based on evolved
N$^3$LO NN potentials and N$^2$LO 3NFs fit to the $^3$H binding energy
and the $^4$He charge radius
(\cite{Hebeler:2010xb}). Theoretical uncertainties are estimated by
the NN (lines)/3N (band) cutoff variations.\label{nm_all}}
\end{figure*}

Three-nucleon forces have also been implemented for neutron-rich
systems.  A frontier in this area is to understand the sequence of
isotopes from proton-rich to the limit of neutron-rich nuclei:~the
neutron dripline. The neutron dripline evolves regularly from light to
medium-mass nuclei except for a striking anomaly in the oxygen
isotopes, where the heaviest isotope, $^{24}$O, is doubly-magic and
anomalously close to stable nuclei (\cite{Janssens:2009fk} and
references therein).  This oxygen anomaly is not reproduced in
shell-model calculations derived from microscopic NN forces (see NN in
Fig.~\ref{oxygen}~(b) and~(c)), only with phenomenological adjustments
(Fig.~\ref{oxygen}~(a)). As shown in Fig.~\ref{oxygen}~(c), chiral
3NFs at N$^2$LO lead to repulsive contributions to the interactions
among excess neutrons that change the location of the neutron dripline
from $^{28}$O to the experimentally observed $^{24}$O
(\cite{Otsuka:2009cs}). This is dominated by the long-range
two-pion-exchange part of 3NFs, as demonstrated in
Fig.~\ref{oxygen}~(b) and~(c) with the single-$\Delta$-excitation
Fujita-Miyazawa 3NF (of the type of Fig.~\ref{fig:FM}). For valence
neutrons, the latter contribution is repulsive, which can be
understood based on the Pauli principle (\cite{Otsuka:2009cs}). This
presents the first microscopic explanation of the oxygen
anomaly. Since the 3NF mechanism is robust and general, these findings
can impact the nucleosynthesis of heavy elements in neutron-rich
environments. The same 3NF contributions have been shown to be key for
the calcium isotopes (\cite{Gallant:2012as}) and for valence-proton
interactions for proton-rich nuclei (\cite{Holt:protonrich}).

Occupying a position between two neutron-rich, doubly-magic isotopes,
$^{22}$O and $^{24}$O, the spectrum of $^{23}$O in Fig.~\ref{23O}
provides a unique test for theory, as it should reflect the features
of both neighbors. In Fig.~\ref{23O}, we observe that 3NF
contributions in extended valence spaces improve the spectrum
considerably (\cite{Holt:2011fj}). With NN forces, the first excited
state is only at $\approx 0.5 \, {\rm MeV}$, well below experiment,
similar to coupled-cluster theory with a N$^3$LO NN potential
(\cite{Hagen:2009mm}). Future studies are needed regarding the
convergence in Fig.~\ref{23O} and the treatment of the center of mass
in such extended valence spaces, as well as to include the continuum
for loosely bound and unbound states (\cite{Michel:2010zq}).

Large-space calculations including the continuum have recently been
carried out for the oxygen and calcium isotopes using coupled-cluster
theory (\cite{Hagen:oxygen,Hagen:calcium}), which lead to a very good
description, especially for excited states and shell structure. These
coupled-cluster calculations include 3NFs as density-dependent
two-body interactions (with adjusted $c_E$ coupling and Fermi momentum
$k_{\rm F}$), developed by \cite{Holt:2009ty} and
\cite{Hebeler:2009iv}, but with different normal-ordering
factors corresponding to two-body forces. The difference
between this approximation and normal-ordering factors for
three-body forces was found to be significant in nuclear
matter calculations (\cite{Hebeler:2010xb}).

Understanding and predicting the formation and evolution of shell
structure from nuclear forces is another key challenge. While the
magic numbers $N=2,8,20$ are generally well understood, $N=28$ is the
first standard magic number that is not reproduced in microscopic
theories with NN forces only (\cite{Caurier:2004gf}). In first studies
for calcium isotopes (\cite{Holt:2010yb}; \cite{Hagen:calcium}),
it was shown that 3NFs are
key to explain the $N=28$ magic number, leading to a high $2^+$
excitation energy and a concentrated magnetic dipole transition
strength in $^{48}$Ca (\cite{vonNeumannCosel:1998plb}).

The calculations of neutron-rich nuclei take into account the
normal-ordered two-body part of 3NFs, which arises from the
interactions of two valence neutrons with a nucleon in the core (see
Fig.~\ref{oxygen}~(d)), which is enhanced by the number of core
nucleons. Moreover, the normal-ordered two-body part can be shown to
dominate over residual three-body interactions based on phase space
arguments for normal Fermi systems (\cite{Friman:2011vm}). The
normal-ordered two-body approximation has been shown to be effective
in coupled-cluster theory (\cite{Hagen:2007ew}) and was
carefully benchmarked for light and medium-mass closed-shell nuclei
(\cite{Roth:2011vt}). In the context of the shell model, residual 3NFs
were recently shown to be small, but amplified with neutron number in
neutron-rich nuclei (\cite{Caesar}). In addition, normal-ordering
techniques have been used to implement the SRG evolution of nuclear
Hamiltonians directly ``in-medium'' in the $A$-body system of interest
(\cite{Tsukiyama:2010rj}), with first results 
including 3NFs (\cite{Hergert:2012}).

Recent developments of chiral EFT and RG transformations for nuclear
forces enable controlled calculations of matter at nuclear densities.
Nuclear matter calculations provide an important benchmark for nuclear
forces, and are used to constrain calculations of heavy nuclei and
matter in astrophysics. The RG evolution to low momenta softens the
short-range tensor components and short-range repulsion of nuclear
forces (\cite{Bogner:2006tw}). This leads to contributions in the
particle-particle channel that are well converged at second order in
the potential, suggesting that perturbative approaches can be used in
place of the Bethe-Brueckner-Goldstone hole-line expansion
(\cite{Bogner:2005sn,Hebeler:2010xb}). In this framework, it is also
possible to estimate the theoretical uncertainties due to neglected
many-body forces and from an incomplete many-body calculation. The
nuclear matter results starting from chiral EFT interactions are shown
in Fig.~\ref{nm_all}. Three-nucleon forces drive saturation, and these
are the first nuclear forces fit only to $A \leqslant 4$ nuclei that
predict realistic saturation properties. For these developments, an
improved treatment of 3NFs as density-dependent two-body interactions
has been key (\cite{Holt:2009ty}; \cite{Hebeler:2009iv}).

The rapid convergence around saturation density in Fig.~\ref{nm_all}
may justify in part the application of in-medium chiral perturbation
theory (\cite{Lutz:1999vc}; \cite{Kaiser:2001jx,Lacour:2009ej}), which
provides an alternative expansion for nuclear densities. In in-medium
chiral perturbation theory, the inclusion of long-range
two-pion-exchange 3NFs from $\Delta$ degrees of freedom also improves
the description of nuclear matter and the convergence
(\cite{Fritsch:2004nx}). In addition, 3NF contributions to the
quasiparticle interactions in nuclear matter have been explored in
\cite{Holt:2011yj}.

The nuclear matter results imply that exchange correlations are
tractable, which opens the door to develop a universal nuclear energy
density functional (UNEDF) for global ground-state predictions based
on microscopic interactions. This is one of the goals of the SciDAC
UNEDF/NUCLEI initiatives. Three-nucleon forces play a key role in
this, including for an improved density matrix expansion based on
chiral EFT interactions (see \cite{Stoitsov:2010ha} and references
therein) and for studies of pairing in nuclei with a
non-empirical pairing functional (\cite{Lesinski:2011rn}).

\begin{table}[t]
\caption{Symmetry energy $E_{\rm sym}$ obtained from neutron matter
calculations with N$^2$LO 3NFs for different $c_1$ and
$c_3$ couplings and based on RG-evolved
N$^3$LO NN forces only (\cite{Hebeler:2010jx}).
\label{tab:Esym}}
\begin{center}
\begin{tabular}{c|c|c}
$\: c_1 \, [{\rm GeV}^{-1}] \:$ & $\: c_3 \, [{\rm GeV}^{-1}] \:$ &
$\: E_{\rm sym} \, [{\rm MeV}] \:$ \\ \hline
\quad $-0.7$ \quad\quad & \quad $-2.2$ \quad\quad & \quad $30.1$
\quad \\
\quad $-1.4$ \quad\quad & \quad $-4.8$ \quad\quad & \quad $34.4$
\quad \\ \hline 
\multicolumn{2}{l|}{NN-only (\cite{Entem:2003ft})} & \quad $26.5$ \quad \\
\multicolumn{2}{l|}{NN-only (\cite{Epelbaum:2004fk})} & \quad $25.6$ \quad
\end{tabular}
\end{center}
\end{table}

For neutron matter, only the long-range two-pion-exchange $c_1$ and
$c_3$ parts of N$^2$LO 3NFs contribute
(\cite{Hebeler:2009iv,Tolos:2007bh}). This has allowed for a detailed
study of the theoretical uncertainties of the neutron matter energy
(\cite{Hebeler:2009iv}). The inclusion of 3NFs leads to an energy per
particle at saturation density $E_n(\rho_0)/N = 16.3 \pm 2.2 \, {\rm
MeV}$, where the uncertainty is dominated by the uncertainty in the
$c_3$ coupling (and to a smaller extent by $c_1$; see the $c_1, c_3$
range in Table~\ref{tab:cis}). Other microscopic calculations lie
within this energy range. The uncertainty of the prediction is again
an estimate of the importance of including N$^3$LO contributions for
neutron and nuclear matter. 
Part of the N$^3$LO 4N forces has been estimated in neutron and
nuclear matter (\cite{Fiorilla:2011sr}), and a first complete N$^3$LO
calculation of neutron matter including NN, 3N and 4N forces has
recently been carried out (\cite{Tews}).

The predicted neutron matter energy also provides constraints for the
symmetry energy (see Table~\ref{tab:Esym}, which demonstrates that the
uncertainty in 3NFs dominates), and predicts the neutron skin
thickness of $^{208}$Pb to $0.17 \pm 0.03 \, {\rm fm}$, in excellent
agreement with a recent determination from the complete electric
dipole response (\cite{Tamii:2011pv}). These developments are
complemented by Auxiliary Field Diffusion Monte Carlo calculations
using a range of phenomenological 3NFs (\cite{Gandolfi:2011xu}) and by
lattice simulations with chiral 3NFs of dilute neutron matter
(\cite{Epelbaum:2009zsa}), which can also enable future benchmarks at
nuclear densities of the perturbative neutron matter calculations

\begin{figure}[t]
\begin{center}
\includegraphics[scale=0.8,clip=]{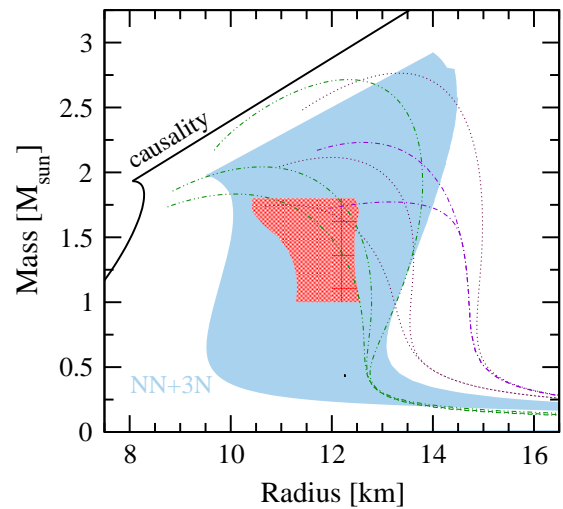}
\end{center}
\caption{(Color online) Mass-radius range for neutron stars based on
chiral EFT NN+3N interactions, combined with a general extrapolation
to high densities (\cite{Hebeler:2010jx}; the light blue/grey region
includes an update for the $1.97 $ neutron star discovered
recently). The predicted range is consistent with astrophysical
modeling of X-ray burst sources (see, e.g., the red/gray shaded 
region from the
\cite{Steiner:2010fz} analysis). For comparison, we also show
equations of state commonly used in supernova simulations (lines
from \cite{Connor:priv}).\label{MvsR}}
\end{figure}

These advances have an important impact on astrophysics. The
microscopic calculations based on chiral EFT interactions constrain
the properties of neutron-rich matter below nuclear densities to a
much higher degree than is reflected in current neutron star modeling
(\cite{Hebeler:2010jx}). Combined with the heaviest $1.97 M_\odot$
neutron star (\cite{Demorest:2010fk}), the neutron matter results
based on chiral NN and 3N interactions constrain the radius of a
typical $1.4 M_\odot$ star to $R \approx 10-14 \, {\rm km}$ ($\pm 15 \%$),
as shown in Fig.~\ref{MvsR}. The predicted radius range is due, in
about equal amounts, to the uncertainty in 3N (and higher-body) forces
and to the extrapolation to high densities. The predicted range is
also consistent with astrophysical results obtained from modeling
X-ray burst sources (see, e.g., \cite{Steiner:2010fz} in
Fig.~\ref{MvsR}). In addition, the comparisons in Fig.~\ref{MvsR}
demonstrate that the constraints resulting from chiral EFT should be
included in equations of state used for simulations of stellar
collapse, neutron stars, and black-hole formation.

%% file: other.tex
\section{Three-body forces and relations to other processes}
\label{sec:other}

Because of gauge symmetries, the same expansion is used to derive
nuclear forces and electroweak operators. Therefore, the couplings of
three-body forces in an EFT determine also electroweak processes. This
is an important consistency test and makes such theories very
predictive.

\begin{figure}[t]
\begin{center}
\includegraphics[scale=0.525,clip=]{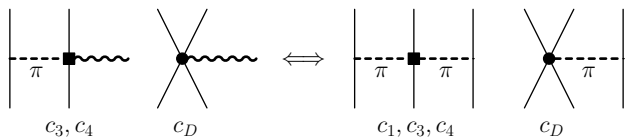}
\end{center}
\caption{Leading two-body axial currents and the corresponding 3NF
contributions in chiral EFT. Solid (dashed) lines indicate nucleons
(pions), and the wavy line represents the axial current.\label{currents}}
\end{figure}

A prime example in chiral EFT are electroweak axial currents, where
pion couplings contribute both to the currents and to nuclear
forces. This is already seen at leading order: $g_A$ determines the
axial one-body current and the one-pion-exchange potential. Two-body
currents, also known as meson-exchange currents, enter at higher
order, just like 3NFs. As shown in Fig.~\ref{currents}, the leading
axial contributions (at order $Q^3$) are due to long-range
one-pion-exchange and short-range parts (\cite{Park:2002yp}), with the
same couplings $c_3, c_4$ and $c_D$ of N$^2$LO 3NFs
(\cite{Gardestig:2006hj,Gazit:2008ma}). Chiral EFT is essential for
this connection, which can be viewed as the two-body analogue of the
Goldberger-Treiman relation.

Two-body currents have also been derived for electromagnetic reactions
(\cite{Pastore:2008ui}; \cite{Kolling:2009iq}; \cite{Pastore:2009is};
\cite{Pastore:2011ip}; \cite{Kolling:2011mt}). Their application to
the few-nucleon system has just started, but based on model
interactions one can expect an interesting sensitivity of many
electromagnetic reactions to two-body currents and 3NFs
(\cite{Golak:2005iy,Bacca:2008tb,Pastore:2012}).  In this colloquium,
we focus on recent developments with electroweak axial currents beyond
light nuclei.

Figure~\ref{currents} demonstrates the unique constraints chiral EFT
provides for two-body axial currents and 3NFs. This relates the
interactions with external probes to the strong-interaction dynamics
in nuclei. In particular, the low-energy coupling $c_D$ that
determines the mid-range one-pion-exchange 3NF can be determined
either from the structure of light nuclei (see
Section~\ref{sec:manybody}); through the two-body axial currents
that enter weak decays such as the $^3$H half-life; or from pion
production in hadronic collisions.\footnote{The low-energy coupling
$c_D$ also enters pion production in NN collisions. However, this
probes significantly higher momenta, because of the produced pion. For
nuclear forces, the determination in pion production may therefore
not be as effective as from the low-momentum kinematics involved in
nuclear structure (\cite{Pandharipande:2005sx}).} This consistency
opens up new avenues of research for weak interactions and fundamental
symmetries.

\begin{figure}[t]
\begin{center}
\includegraphics[scale=0.325,clip=]{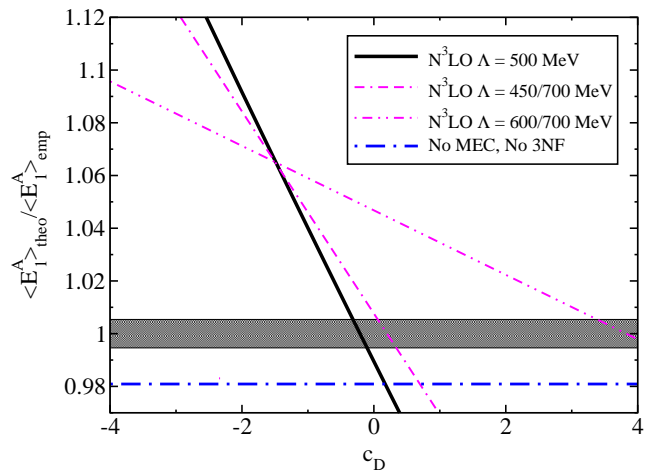}
\end{center}
\caption{(Color online) The ratio $\langle E_1^A \rangle_{\rm theo}/
\langle E_1^A \rangle_{\rm emp}$ that determines the $^3$H half-life
as a function of the low-energy coupling $c_D$, which relates the
leading two-body axial currents and 3NFs (see Fig.~\ref{currents}).
The empirical range is given by the horizontal band.
Results are shown based on different N$^3$LO NN potentials and
including N$^2$LO 3NFs and consistent two-body axial currents.
For comparison, the result without 3NFs and without two-body 
currents (No MEC, No 3NF) is given. For details see 
\cite{Gazit:2008ma}.\label{triton}}
\end{figure}

Although the importance of two-body currents is known from
phenomenological studies, for weak processes, chiral currents and the
consistency with nuclear forces have only been explored in light
nuclei (\cite{Park:2002yp}; \cite{Gazit:2008ma};
\cite{Kubodera:2010qx}).  Figure~\ref{triton} shows the dependence of
the $^3$H half-life on the low-energy coupling $c_D$, which is
included both in the leading 3NFs and two-body axial currents. Without
3NFs and without two-body currents, the experimental $^3$H half-life
is not reproduced. A dependence on the different N$^3$LO NN potentials
is expected, because the leading 3NFs and two-body axial currents are
only order $Q^3$. As for 3NFs, the next order two-body axial currents
are predicted in chiral EFT, without free parameters, which enables
systematic improvements of beta-decay studies and predictions. The
chiral EFT currents determined from the $^3$H half-life have recently
been applied to the beta decay of the two-neutron halo nucleus $^6$He
(\cite{Vaintraub:2009mm}), however using a phenomenological potential
model not based on chiral EFT, where the decay rate is satisfactorily
reproduced. These theoretical studies are complemented by precision
measurements (see, e.g., the recent result for the $^6$He half-life
(\cite{Knecht:2011ir})).

Surprisingly, key aspects of well-known beta decays in medium-mass
nuclei remain a puzzle. In particular, when calculations of
Gamow-Teller (GT) transitions of the spin-isospin operator $g_A {\bm
\sigma} \tau^{\pm}$ are confronted with experiment (this is the most
significant operator for beta decays and for electron-capture
processes), some degree of renormalization, or ``quenching'' $q$, of
the axial coupling $g_A^{\rm eff} = q g_A$ is needed. Compared to the
single-nucleon value $g_A=1.2695(29)$, the GT term seems to be weaker
in nuclei. This was first conjectured in studies of beta-decay rates,
with a typical $q \approx 0.75$ in shell-model calculations
(\cite{Wildenthal:1983zz}; \cite{MartinezPinedo:1996vz}) and other
many-body approaches (\cite{Bender:2001up,Rodriguez:2010mn}). In view
of the significant effect on weak reaction rates, it is no surprise
that this suppression has been the target of many theoretical works
(see the discussion in \cite{Vaintraub:2009mm}).

Recent studies of GT transitions in medium-mass nuclei with chiral EFT
currents provide new insights and opportunities to this puzzle
(\cite{Menendez:2011qq}). Compared to light nuclei, the contributions
of chiral two-body currents are amplified in medium-mass nuclei
because of the larger nucleon momenta. Using a normal-ordering
approximation for two-body currents to create a density-dependent
operator, it was shown that the leading two-body axial currents
contribute only to the GT operator (up to a small tensor-like
correction) and that a quenching of low-momentum-transfer GT
transitions is predicted based on the long-range parts of two-body
currents. This demonstrates that chiral two-body currents naturally
contribute to the quenching of GT transitions. A reduction of $g_A$ in
the currents is also expected considering chiral 3NFs as
density-dependent two-body interactions
(\cite{Holt:2009uk,Holt:2009ty}).  The long-range one-body
contributions from two-body currents are in part due to Delta-hole
pairs, but it remains an open problem how much of the quenching of
$g_A$ is due to two-body currents and how much due to polarization effects.

\begin{figure}[t]
\begin{center}
\includegraphics[scale=0.375,clip=]{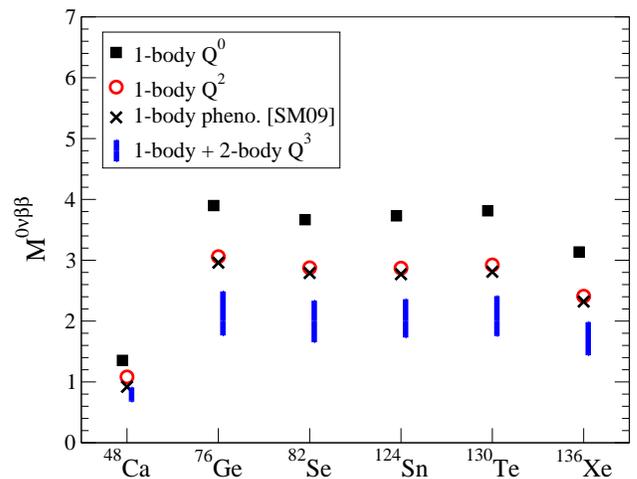}
\end{center}
\caption{(Color online) Nuclear matrix elements $M^{0\nu\beta\beta}$
for neutrino-less double-beta decay of different nuclei. Results
are shown based on chiral EFT currents at successive orders, including
one-body currents at orders $Q^0$ and $Q^2$, and the predicted 
long-range parts of two-body currents at order $Q^3$
(\cite{Menendez:2011qq}; for a discussion of the short-range
contributions, see this reference). For comparison, we also show 
shell-model results (SM09) of \cite{Menendez:2008jp} based on
phenomenological one-body currents only.\label{nme}}
\end{figure}

Neutrinoless double-beta decay presents a fundamental test of the
nature of the neutrino, of lepton number, and the neutrino mass scale
and hierarchy (\cite{Elliott:2002xe}; \cite{Avignone:2007fu}). A
pivotal input for the ongoing and planned experimental searches are
the nuclear matrix elements that incorporate the structure of the
parent and daughter nuclei and of the decay mechanism. Compared to
standard beta decays, neutrinoless double-beta decay probes different
momentum transfers $Q \approx 100 \, {\rm MeV} \sim m_\pi$
(\cite{Simkovic:2007vu}; \cite{Menendez:2011qq}). Therefore, the
impact of two-body currents is unclear and renormalization effects can
differ from the suppression of $g_A$ in medium-mass nuclei. Chiral EFT
predicts the momentum-transfer dependence of two-body currents, which
varies on the order of the pion mass due to the one-pion-exchange part
in Fig.~\ref{currents}. The first calculation of the neutrinoless
double-beta decay operator based on chiral EFT currents at successive
order is shown in Fig.~\ref{nme}. This demonstrates that the
contributions from two-body currents are significant and should be
included in all calculations. It also shows how chiral EFT can provide
important input and theoretical uncertainties for fundamental symmetry tests
with nuclei. Recently, chiral EFT currents have also been applied to
calculate the structure factor for spin-dependent weakly interacting
massive particle (WIMP) scattering off nuclei, needed for direct
dark matter detection (\cite{WIMP}).

%% file: outlook.tex
\section{Outlook and future opportunities}
\label{outlook}

\begin{figure*}[t!]
\begin{center}
\begin{tabular}{l|ccc}
& \: {\large pionless} & {\large chiral} & \:\: {\large chiral$+\Delta$}
\\[5pt] \hline \\[-2pt]
\large LO & \quad
\parbox{1cm}{\includegraphics[height=1cm]{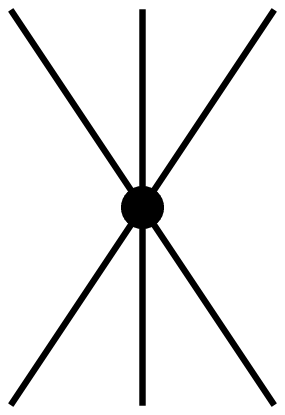}} 
\quad & \Large --- & \:\: \Large --- \cr \cr
\large NLO & \: \Large --- & \Large --- & \quad
\parbox{1cm}{\includegraphics[height=1cm]{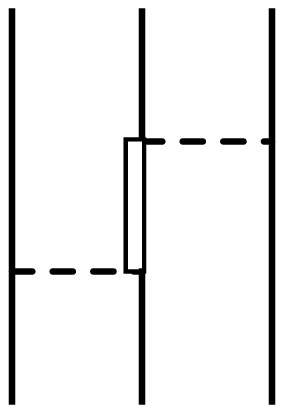}} 
\cr \cr 
\large N$^{2}$LO \hspace*{1pt} & \quad
\parbox{1cm}{\includegraphics[height=1cm]{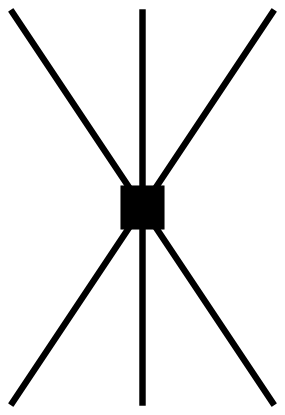}}
\quad & \quad
\parbox{1cm}{\includegraphics[height=1cm]{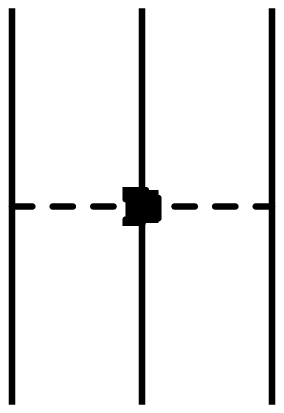}} 
\parbox{1cm}{\includegraphics[height=1cm]{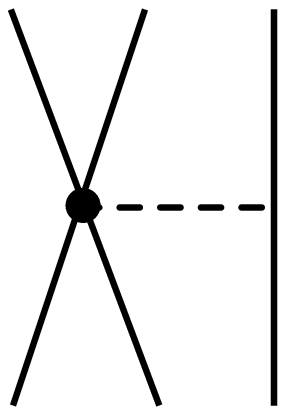}}
\parbox{1cm}{\includegraphics[height=1cm]{tnf-contact.eps}}
\quad & \quad
\parbox{1cm}{\includegraphics[height=1cm]{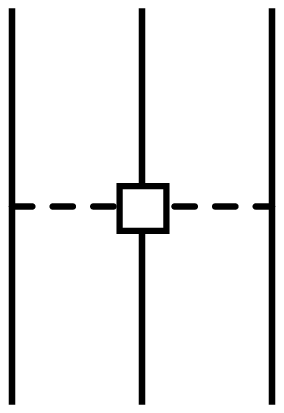}}
\parbox{1cm}{\includegraphics[height=1cm]{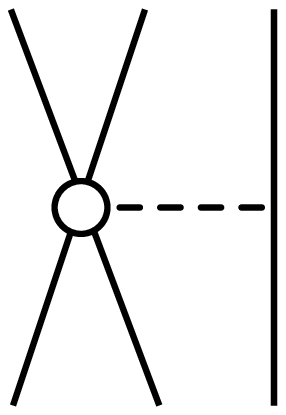}}
\parbox{1cm}{\includegraphics[height=1cm]{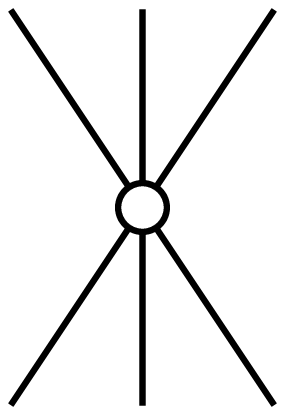}}
\end{tabular}
\end{center}
\caption{Order of 3NF contributions in pionless and chiral EFT and in
EFT with explicit $\Delta$ degrees of freedom (chiral$+\Delta$). Open vertices in the last 
column indicate the differences of the low-energy constants in chiral and chiral$+\Delta$ EFT.
\label{fig:3nfsum}}
\end{figure*}

In this colloquium, we have highlighted the importance of three-body
forces in nuclear physics and related areas. Here, we give an outlook
with a focus on future opportunities and challenges. Our discussion is
guided by Fig.~\ref{fig:3nfsum} which summarizes the leading 3NFs in
different EFTs and shows the order in the expansion at which they
enter.

In pionless EFT for systems with large two-body scattering lengths
discussed in Section~\ref{sec:univ}, three-body forces contribute
already at leading order because of the Efimov effect. If observables
are considered at fixed scattering lengths, subleading three-body
forces are suppressed by two orders and enter only at N$^2$LO. Some
higher-order calculations of few-nucleon observables exist but much
remains to be investigated in this sector. Particularly interesting
are the application of pionless EFT to halo nuclei and low-energy
electroweak reactions. Halo nuclei are the most promising candidates
for observing Efimov physics in nuclei, while precise calculations of
low-energy reactions are relevant for nuclear astrophysics and
neutrino physics. In particular, 3NFs play a prominent role in
two-neutron halo nuclei and larger halo systems.  Pionless EFT also
predicts universal three-body correlations that can be explored in
nuclear reactions in this regime and to test the consistency of
different theoretical calculations (similar to the Tjon line/band).

In chiral EFT discussed in Sections~\ref{sec:chpt},~\ref{sec:manybody}
and~\ref{sec:other}, 3NFs are suppressed compared to NN interactions.
This explains the phenomenological success of weaker three-body forces
of the Fujita-Miyazawa type. As summarized in Fig.~\ref{fig:3nfsum},
3NFs enter at N$^2$LO, and their relative contributions to observables
can be understood based on the power counting. Because the operator
structure of the leading 3NFs is strongly constrained, a global
analysis of few-body scattering and bound-state data with theoretical
uncertainties appears feasible in the framework of chiral EFT. This
would allow for a determination of the long-range $c_i$ couplings in
the three-body sector. In addition, a consistent determination of two-
and three-body forces from such an analysis may help to resolve the
$A_y$ puzzle in few-body scattering.

For applications of chiral EFT interactions to nuclear structure, 3NFs
play a central role, as discussed for light and medium-mass nuclei and
for nuclear matter. For these many-body calculations, the RG/SRG evolution
leads to greatly improved convergence. A consistent evolution of
chiral 3NFs has been achieved in a harmonic-oscillator basis and
recently in momentum space. Important open problems are an
understanding of the 3NFs induced by the SRG and to control
higher-body forces, which is necessary for the desired accuracy in
nuclear structure.

If $\Delta(1232)$ degrees of freedom are included, part of the physics
contained in the low-energy constants in chiral EFT is made explicit
in lower orders. As a consequence, a 3NF of the Fujita-Miyazawa type
appears already at NLO as shown in Fig.~\ref{fig:3nfsum}. Improved
convergence of the chiral expansion with explicit $\Delta$ degrees of
freedom is expected, but a full analysis of few-nucleon data remains
to be carried out. In addition, a chiral EFT with explicit $\Delta$'s
would naturally explain why the contributions from the long-range
two-pion-exchange parts of 3NFs dominate over the shorter-range parts
in applications to neutron-rich nuclei and nuclear matter.

Three-nucleon forces are a frontier in the physics of nuclei that
connects the systematic development of nuclear forces in chiral EFT
with the exploration of neutron-rich nuclei at rare isotope beam
facilities. The subleading 3NFs at N$^3$LO are predicted in chiral
EFT, without free parameters, as is the case for N$^3$LO 4N forces.
In many present calculations, the uncertainty of the leading 3NFs
likely dominates the theoretical uncertainties of the predicted
observables. The derivation of N$^3$LO 3NFs has only been completed
recently, and no calculation exists with N$^3$LO 3N or 4N forces
beyond few-body systems. Therefore, there is a window of opportunity
to make key discoveries and predictions. In addition to advancing
microscopic calculations with 3NFs to larger and neutron-rich nuclei,
an important problem is to understand the impact of 3NFs on global
nuclear structure predictions, e.g., for key regions in the r-process
path where systematic theoretical predictions of extreme nuclei, often
not accessible in the laboratory, are needed.

Electroweak interaction processes are unique probes of the physics of
nuclei and fundamental symmetries, and play a central role in
astrophysics. Chiral EFT provides a systematic basis for nuclear
forces and consistent electroweak currents, where pion couplings
contribute both to electroweak currents and to 3NFs. This opens up new
opportunities for precise nuclear structure calculations with
theoretical uncertainties that are needed for fundamental symmetry
tests with beta decays and weak transitions, including the key nuclear
matrix elements for neutrinoless double-beta decay.

In principle, it is possible to calculate nuclear properties directly
from the QCD Lagrangian. In Lattice QCD, the QCD path integral is
evaluated in a discretized Euclidean space-time using Monte Carlo
simulations. 
This approach is based on a nonperturbative formulation of QCD but
requires a large numerical effort. However, high statistics Lattice
QCD simulations of two- and three-nucleon systems are now within reach
(\cite{Beane:2010em}) and the calculation of few-nucleon systems
appears feasible in the intermediate future. A milestone for nuclear
forces is the prediction of three-neutron properties in a box.  This
will provide unique access to the isospin $T=3/2$ component of 3NFs,
which is not probed in nucleon-deuteron scattering. Moreover, Lattice
QCD results can also be used to constrain couplings in chiral 3NFs. A
first step in this direction was recently carried out by
\cite{Doi:2011gq}.

\begin{acknowledgments}
We thank R.\ J.\ Furnstahl, U.-G.\ Mei{\ss}ner, D.\ R.\ Phillips and
W.\ Weise for comments on the manuscript. Many discussions with our
collaborators on the topics of this colloquium are gratefully
acknowledged. This work was supported by the DFG through SFB/TR 16
and SFB 634, by the BMBF under contracts No.~06BN9006 and
No.~06DA70471, by the JSC, J\"ulich, by the Helmholtz Alliance
Program of the Helmholtz Association, contract HA216/EMMI ``Extremes
of Density and Temperature: Cosmic Matter in the Laboratory'', and
by the ERC grant No.~307986 STRONGINT.
\end{acknowledgments}